\documentclass[aps,prb,reprint,superscriptaddress,nofootinbib]{revtex4-1}
\usepackage{amsmath,amssymb}
\usepackage{tikz}
\usepackage{graphicx}
\usepackage{epstopdf}
\usepackage{bm}
\usepackage{bbold}
\usepackage{color}
\usepackage{amsmath}
\usepackage{amsfonts}
\usepackage{amssymb}
\usepackage{multirow}
\usepackage{float}
\usepackage{booktabs}
\usepackage{hyperref}
\usepackage{array}

\newcommand{\ga}
{\alpha}
\newcommand{\gb}
{\beta}

\newcommand*\circled[1]{\tikz[baseline=(char.base)]{\node[shape=circle,draw,inner sep=1pt] (char) {#1};}}

\begin{document}

\title{Unified low-energy effective Hamiltonian and the band topology of $p$-block square-net layer derivatives}

\author{S. I. Hyun}
\affiliation{Department of Chemistry, Pohang University of Science and Technology, Pohang 37673, Korea}

\author{Inho Lee}
\affiliation{Department of Chemistry, Pohang University of Science and Technology, Pohang 37673, Korea}

\author{Geunsik Lee}
\email{gslee@unist.ac.kr}
\affiliation{Department of Chemistry, School of Natural Science, Ulsan National Institute of Science and Technology, Ulsan 44919, Korea}

\author{J. H. Shim}
\email{jhshim@postech.ac.kr}
\affiliation{Department of Chemistry, Pohang University of Science and Technology, Pohang 37673, Korea}
\affiliation{Department of Physics, Pohang University of Science and Technology, Pohang 37673, Korea}
\date{\today}

\begin{abstract}
In recent years, low-dimensional materials with tetragonal \textit{P4/nmm} (orthorhombic \textit{Pnma}) space group having square-net (chain-like) substructure of $p$-block elements have been studied extensively. By using a first-principles calculation and a two-sites $\otimes$ two-orbitals tight-binding model, we construct the unified low-energy effective Hamiltonian and the $\mathbb{Z}_{2}$ topological phase diagram for such materials with different filling factors. Near the chemical potential, we show that the staggered arrangement of ions at 2c (4c) site yields the virtual hopping that have the same form with the second nearest-neighbor hopping between the square-net (chain-like) ions. 
We show that this hybridization and low-symmetry of the chain-like structure protects the quantum spin Hall insulator phase. Finally, the second order spin-orbit coupling on top of the atomic spin-orbit coupling is considered to clarify the origin of the non-zero Berry phase signals reported in recent quantum oscillation experiments.
\end{abstract}

\maketitle

\section{Introduction}
\begin{figure}[bp]
	\centering
	\includegraphics[width=1.0\linewidth]{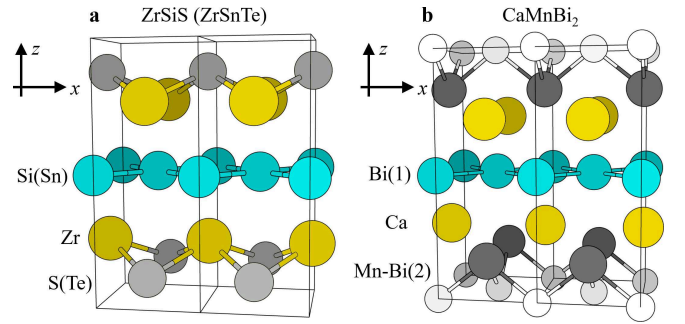}
	\caption{(Color Online) Crystal structure of \textbf{a} ZrSiS, \textbf{b} CaMnBi$_{2}$, where square-net ions are denoted as light blue and the staggeredly-stacked ions at 2c site are denoted as yellow in both cases. Bi(1) and Bi(2) in \textbf{b} indicate the first and second types of Bi atoms, respectively.}
	\label{fig:structure}
\end{figure}

Since Young and Kane \cite*{Young2015} proved the existence of the symmetry-protected Dirac electron in quasi 2-dimensional system with nonsymmorphic space group (S.G.), their work has been generalized and extended to the magnetic material or 3-dimensional symmetries\cite*{Yang2017,Park2017,Wang2017}. In the experimental side, accessible nonsymmorphic degeneracy and the existence of surface state have been studied by using angle-resolved photoemission spectroscopy (ARPES), and Dirac-like behavior, i.e., non-zero Berry phase has been revealed by using Shubnikov-de Haas (SdH) and other quantum oscillation experiments\cite*{Neupane2016,Hosen2017,Hu2016,Ali2016,Schoop2015,Topp2016,Topp2017,Lou2016,Liu2016}. As shown in Fig.\hyperref[fig:structure]{\ref*{fig:structure}}, there are variety of materials, which are featured commonly by a square lattice layer of the $p$-block element. Depending on the mother-material, we divide the materials into two families. One is based on the Zr$AQ$ ($A$=Si, Sn, $Q$=O, S, Se,Te) and the other is based on $AE$Mn$B$$_{2}$ ($AE$=Sr, Ca, $B$=Bi, Sb). Although they share similar substructure, these families are classified according to the filling factor of the substructure and the hybridization strength\cite*{Xu2015}. Zr$AQ$-family with \textit{P4/nmm} (S.G. 129) has a square-net substructure constructed with carbon group (group 14) atoms of $A$=Si or Sn.  Meanwhile, $AE$Mn$B$$_{2}$-family has the square-net \textit{P4/nmm} or the chain-like \textit{Pnma} (S.G. 62) substructure of pnictogens (group 15) $B$=Bi or Sb.

As mentioned before, these $p$-electrons exhibit an intriguing electronic structures driven by \textcolor{black}{the crystal} symmetry, and there have been extensive amount of experimental investigations on the both families. However, no unified description of $p$-orbital band is available to give an insight of the role of chemical and structural details, and thorough study of all possible band topologies are needed. So we devote this paper (i) to unify the low-energy effective Hamiltonian for various square-net and chain-like substructure of $p$-block elements and (ii) to reveal the topological phase diagram and the source of the non-zero Berry phase.

\begin{figure*}[htbp]
	\centering
	\includegraphics[width=1.0\linewidth]{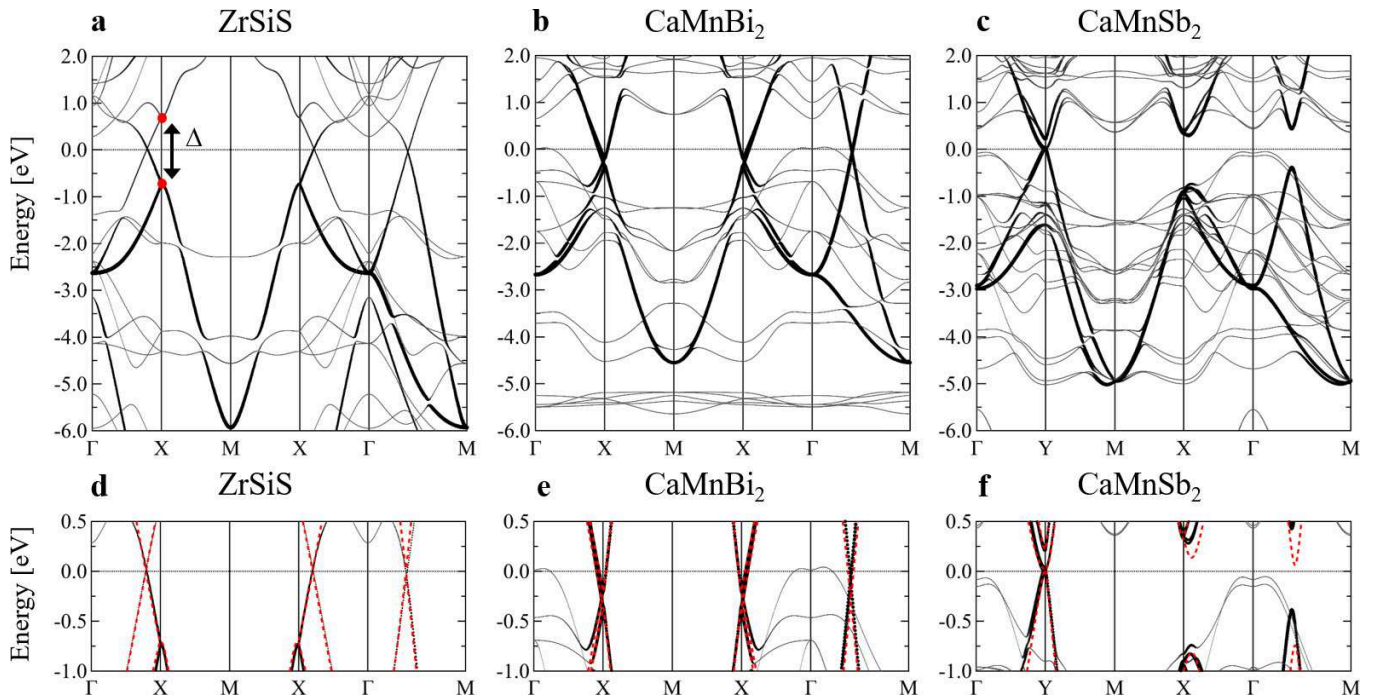}
	\caption{(Color Online) Band structure of \textbf{a} ZrSiS, \textbf{b} CaMnBi$_{2}$, and \textbf{c} CaMnSb$_{2}$ in large energy-window. Low-energy band structures are given in \textbf{d-f} in the same order. $p_{x,y}$ states from the square-net or chain-like $p$-block elements are presented in heavier dots. In \textbf{d-f}, TB dispersions are given in red dashed lines. }
	\label{fig:dispersion}
\end{figure*}

Hence, we organize this paper as follows. Section \hyperref[sec:dft]{\ref*{sec:dft}} presents the density functional theory (DFT) band structures of selected materials. We select ZrSiS and CaMnBi$_{2}$ for square-net system and CaMnSb$_{2}$ for chain-like system.  In Sec. \hyperref[sec:tb]{\ref*{sec:tb}}, we build the effective tight binding (TB) Hamiltonian for both Zr$AQ$-family and $AE$Mn$B$$_{2}$-family by employing the second order perturbation theory on the interlayer orbital hybridization by Zr or Ca and treating the chain-formation parameters as the Peierls-distortion\cite*{Su1979, Su1980, Heeger1988, Peierls2001}. In Sec. \hyperref[sec:phase]{\ref*{sec:phase}}, we present topological phase diagram in the presence of the chain-formation parameters based on the band connectivity of the effective Hamiltonian. In Sec. \hyperref[sec:berry]{\ref*{sec:berry}}, the connection between the topology, i.e., band connectivity and the low-energy band structures is explained by investigating the Kane-Mele type second order spin-orbit coupling (SOC). Finally, we briefly discuss the importance of electron-hole asymmetry in Sec.\hyperref[sec:discussion]{\ref*{sec:discussion}} and conclude our paper in Sec. \hyperref[sec:conclusion]{\ref*{sec:conclusion}}.


\section{DFT Calculation}\label{sec:dft}

Electronic structures are calculated by using the full-potential linearized augmented plane-wave method implemented in the WIEN2k package. \cite*{P.BlahaK.SchwarzG.K.H.Madsen2001} The generalized gradient approximation by Perdew-Burke-Ernzerhof (PBE-GGA) is used for the exchange-correlation potential. The modified Becke Johnson Potential (mBJ) \cite*{Becke2006, Tran2009, Koller2011} is employed to minimize the shortcomings of PBE-GGA in the misjudgment of the crystal field splitting. The PBE-GGA+mBJ potential using the mBJ exchange potential plus the GGA is known to calculate crystal field precisely\cite*{Li2014,Kang2016}. For the charge self-consistent calculation, the $k$-mesh $(k_{x},k_{y},k_{z})$ used in the first Brillouin zone is (22, 22, 9) for ZrSiS, (23, 23, 9) for CaMnBi$_{2}$, and (29, 29, 5) for CaMnSb$_{2}$ with RKmax of 7.00. 

Fig. \hyperref[fig:dispersion]{\ref*{fig:dispersion}} shows the DFT band structures of ZrSiS, CaMnBi$_{2}$, and CaMnSb$_{2}$ without SOC. Red-dashed lines are the calculated eigenvalues of the non-relativistic TB Hamiltonian introduced in section \hyperref[sec:tb]{\ref*{sec:tb}}. As indicated by the size of the points, Si-$p_{x}$ and $p_{y}$ orbitals of ZrSiS and Bi(Sb)-$p_{x,y}$ orbitals of CaMnBi(Sb)$_{2}$ show the similar dispersion especially in the valence bands. In the case of ZrSiS, valence states are mainly composed of Si-$p_{x,y}$ states, whereas the conduction bands are Zr-$d$ states. In the case of CaMnBi(Sb)$_{2}$, however, low-energy band structures have only Bi(Sb)-$p_{x,y}$\cite*{Xu2015,Lee2013} characters in both valence and conduction bands.

\textcolor{black}{We integrated $p_{x,y}$ density of states (DOS) to calculate the occupancy of such states. Lower limit of the integration is the energy of the band minimum located at M point and the upper limit is the chemical potential denoted as 0.0 [eV] in Fig. \hyperref[fig:dispersion]{\ref*{fig:dispersion}}.} Numerical integration of DOS shows that the occupancy of Bi(Sb)-$p_{x,y}$ states is very close to half-filled where that of Si-$p_{x,y}$ is close to quarter-filled. There is, however, a striking similarity in the low-energy structure of ZrSiS and CaMnBi$_{2}$. 
Low-energy structures of CaMnBi$_{2}$ is very similar to the pristine half-filled square-net array since the virtual hopping between the empty Ca-3$d$ and Bi-6$p$ is minimal due to the energy difference around 5 eV\cite*{Lee2013}. So, CaMnBi$_{2}$ is described as the nodal-line semimetal by the zone-folding. \cite*{Lee2013} On the other hands, ZrSiS has substantial contribution of Zr-4$d$ near the chemical potential and still shows the similar low-energy structure with CaMnBi$_{2}$. We will discuss this apparent similarity in section \hyperref[sec:tb]{\ref*{sec:tb}} by employing the second order perturbation theory. And by considering Kane-Mele type SOC in section \hyperref[sec:berry]{\ref*{sec:berry}} to study the SOC-related low-energy gap along the nodal-line.

It is known that when the ionic size of the square-net is reduced, chain-like distortion occurs in the square-net substructure\cite*{Anderson1990}. If the Bi in CaMnBi$_{2}$ are changed to Sb, the symmetry is reduced from \textit{P4/nmm} (S.G. 129) to \textit{Pnma} (S.G. 62). \textcolor{black}{In the case of CaMnSb$_{2}$, Sb atoms are displaced along the $y$ direction to form zigzag-shaped Sb chains along the $x$ direction as shown in Fig. \hyperref[fig:variation]{\ref*{fig:variation} \textbf{b}}}. Throughout this paper, we will use the chain-formation parameters for the description of its contribution to the low-energy structure. Since there are huge differences in the low-energy structure by the distortion, we inspect how the band structure changes against the distortion by using TB analysis. DFT shows nodal-line semimetal phase in both ZrSiS and CaMnBi$_{2}$. Even with the difference in the hybridizing $d$-orbital energy level, there seems to be the apparent similarity in two cases. In the following section we build the TB model to explain the evolution of low-energy electronic structure with the different filling factor.

\section{TB Analysis}\label{sec:tb}

\subsection{Square-net layer}\label{subsec:square}

\begin{figure}[bp]
	\centering
	\includegraphics[width=1.0\linewidth]{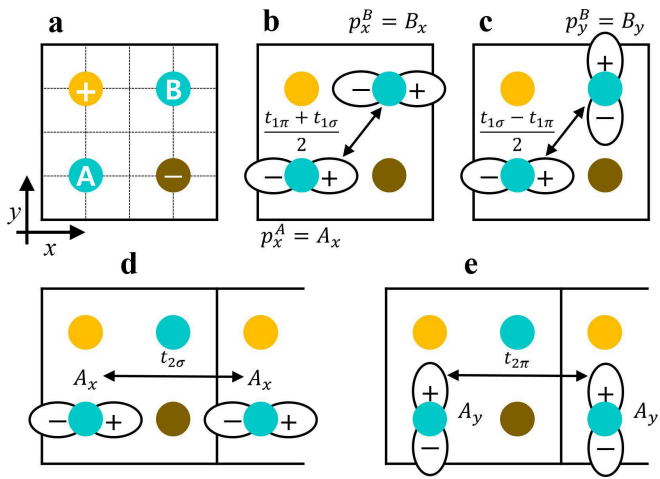}
	\caption{(Color online) Schematic illustration of square-net $p$-block ions (A, B) and staggered $d$-block ions at 2c site ($\pm$) in the unit cell. Abbreviations for the local orbitals and their nearest neighbor hopping process between the same (different) orbitals are given in \textbf{b} (\textbf{c}). Likewise, \textbf{d}(\textbf{e}) illustrates the $\sigma$- ($\pi-$) type next-nearest neighbor (the second nearest neighbor) hopping process between the same orbitals.} 
	\label{fig:hopping}
\end{figure}

Previous \textit{ab initio} calculations and TB analysis on CaMnBi$_{2}$ indicate that the square-net substructure is sufficient to describe the low-energy band structure of the system\cite*{Lee2013}. 
Here, we generalize the model to consider chain-type as well as Zr\textit{AQ}-family lattices. Because of the unit cell (UC) doubling, there are two atomic sites A $(a/4,a/4,0)$ and B $(3a/4,3a/4,0)$ in the primitive UC of the square-net layer with the lattice constant $a$ as shown in Fig.\hyperref[fig:hopping]{\ref*{fig:hopping} \textbf{a}}. The square-net layer is formed by Si or Sn in the case of Zr\textit{AQ}-family and Bi or Sb in the case of \textit{AE}Mn\textit{B}$_{2}$-family.
Nearby the layer, the 2c sites $(a/4,3a/4,+z_{c})$ and $(3a/4,a/4,-z_{c})$ are occupied by Zr in the case of Zr\textit{AQ}-family and Sr or Ca in the case of \textit{AE}Mn\textit{B}$_{2}$-family. Ions sitting at 2c sites are represented in yellow and brown dots in Fig.\hyperref[fig:hopping]{\ref*{fig:hopping} \textbf{a}}.  

First, we consider the square-net layer without including the ions at 2c site. We build the effective TB model by choosing the innate 4-component basis of the doubled UC, that allows to simplify the representation of the parity operator.\cite*{Maier2013,Bena2009,Kochan2017} 
It includes only $p_x$, $p_y$ orbitals centred at A and B sublattice atoms, which we denote as $\Psi_{\bm{R}}\equiv\left[ A_x(\bm{R}), A_y(\bm{R}), B_x(\bm{R}), B_y(\bm{R}) \right]^{\textrm{T}}$ at each square-lattice point $\bm{R}$. Thus the effective Hamiltonian reads,
\begin{equation}
H=\sum_{\bm{RR'},\alpha\beta}\Psi_{\bm{R}\alpha}^{\dagger}V_{\bm{RR'}}^{\alpha\beta}\Psi_{\bm{R'}\beta}.
\end{equation}
The operator $\Psi_{\bm{R}\alpha}^{\dagger}$ denotes the creation of an 
electron at $\alpha$-th component of $\Psi_{\bm{R}}$, and $V_{\bm{RR'}}^{\alpha
\beta}$ is the Slater-Koster (SK) parameter between two components, $\alpha
$ of $\Psi_{\bm{R}}$ and $\beta$ of $\Psi_{\bm{R'}}$.
In terms of the Bloch orbitals, $\Psi_{\bm{k}}=(1/\sqrt{N})\sum_{\bm{R}}\exp\left[-i \bm{k}\cdot\bm{R}\right]\Psi_{\bm{R}}$, in the momentum space, the Hamiltonian  becomes
\begin{equation}
H=\sum_{\bm{k},\alpha\beta}\Psi_{\bm{k}\alpha}^{\dagger} V_{\bm{k}}^{\alpha\beta}\Psi_{\bm{k}\beta}.
\end{equation}

Due to the hermiticity, the Hamiltonian (density) can be spanned by the Kronecker product of the two Pauli matrices. We will assign Pauli matrices $\tau_{i}$ and $\sigma_{i}$ for site- and orbital-space, respectively where $\tau_{0}=\sigma_{0}=\mathbb{1}_{2\times2}$. The Hamiltonian density at each $\bm{k}$ is thus 

\begin{equation}
h_{\bm{k}}=\sum_{i,j=0,1,2,3} d^{ij}_{\bm{k}}\tau_{i}\otimes\sigma_{j}.
\end{equation}

In our TB model, the hopping up to the nearest neighbors is considered. Hopping processes and the form of basis orbitals are presented in Fig.\hyperref[fig:hopping]{\ref*{fig:hopping} \textbf{b-d}}. In our convention of defining the Bloch orbital, denoted as the UC-basis, an additional phase appears, being different from the effective Hamiltonian derived in the site-basis. \cite*{Lee2013} The effective Hamiltonian density for the pristine square-net layer reads, 

\begin{equation}
h_{\bm{k}}^{(0)} =\left(\begin{array}{cc} 
\varepsilon_{p}\sigma_{0} & e^{i \Lambda_{\bm{k}}} V_{pp}\\ 
e^{-i\Lambda_{\bm{k}}} V_{pp} & \varepsilon_{p}\sigma_{0}  \\ 
\end{array}\right),\quad \Lambda_{\bm{k}}=\frac{a k_{x}+a k_{y}}{2}.
\end{equation}

Here, $V_{pp}=t^{10} \cos\frac{a k_{x}}{2}\cos\frac{a k_{y}}{2}\sigma_{0}+t^{11} \sin\frac{a k_{x}}{2}\sin\frac{a k_{y}}{2}\sigma_{1}$ and $\varepsilon_{p}$ denotes the energy level of $p_{x}$ and $p_{y}$ orbitals. In terms of SK parameters, $t^{10}=2(t_{1\pi}+t_{1\sigma}), t^{11}=2(t_{1\pi}-t_{1\sigma})$. 
The phase $\Lambda_{\bm{k}}$ will not appear, if a relative phase factor $\exp[-i\bm{k}\cdot(a/2,a/2,0)]$ between A and B atoms is taken into account like the case in the site-basis.  
In other words, the phase $\Lambda_{\bm{k}}$ may or may not appear in $h_{\bm{k}}$ according to the convention used in defining the center of the Bloch orbital. We write $h_{\bm{k}}^{\textrm{UC}}$ for the former case, $h_{\bm{k}}^{\textrm{site}}$ for the latter case. Superscript UC and site denote the center of Bloch orbital.\cite*{Bena2009}  Between these two, there exists a unitary transformation $h_{\bm{k}}^{\textrm{UC}}=U_{\bm{k}} h_{\bm{k}}^{\textrm{site}}U_{\bm{k}}^{\dagger}$. The unitary matrix is given by $U_{\bm{k}}=\exp[i \Lambda_{\bm{k}} \tau_{3}/2]$. 
In case of Bi-lattice, the nodal-line semimetal phase can be obtained since the energy of the square-net atoms are located exactly at the chemical potential where the band-folding occurs. In the Si-lattice, however, there are two less electrons than Bi-lattice. Therefore, it is essential to add the interaction by the 2c site ions to describe the similar low-energy structure in both materials.

\subsection{Perturbation by 2c sites}\label{subsec:hybridization}

In this section, the apparent agreement in the low-energy band structures between the ZrSiS and CaMnBi$_{2}$ is investigated under the second order perturbation. It is known that the TB model of single CaBi-layer can describe the DFT band structure of CaMnBi$_{2}$.\cite*{Lee2013} However, the dimension of the previously reported Hamiltonian is large since it contains all the information about $p$ and $d$ states. Hence, we will treat the hybridization between square-net $p$-orbital and 2c site $d$-orbitals as the perturbation to derive the effective low-energy Hamiltonian in terms of modified $p$-orbitals of the square-net system. Model system has $p_{x,y}$ orbitals located at $(a/4,a/4,0)$ and $(3a/4,3a/4,0)$ marked as A and B in Fig.\hyperref[fig:hopping]{\ref*{fig:hopping} \textbf{a}} and $d$-orbitals located at 2c sites $(a/4,3a/4,+z_{2c})$ and $(3a/4,a/4,-z_{2c})$ marked as $\pm$ in the same figure. Unperturbed Hamiltonian of the model system $H_{0}$ is given by previously derived $h_{pp}^{0}$ and energy of $d$-orbitals. To simplify the problem, crystal splitting in $d$-states is neglected. $pd$-hybridization matrix $V$ is calculated using SK parametrization with $z_{2c}=a/2$. Note that specific value of $z_{2c}$ does not change the functional form of the final result.

\begin{equation}\label{eq:hybridization}
H_{0}=\left(\begin{array}{cc} 
h_{\bm{k}}^{(0)} & 0\\ 
0 & \varepsilon_{d}\times \mathbb{1}_{10\times10} \\ 
\end{array}\right),
V=\left(\begin{array}{cc} 
0 & V_{pd}\\ 
V_{dp} & 0 \\ 
\end{array}\right).
\end{equation}


Within the second order perturbation theory on $V_{pd}$, the $p$-orbital projected Hamiltonian density $h_{\bm{k}}$ \textcolor{black}{near the chemical potential $(\mu)$} can be given as 

\begin{equation}\label{eq:hybridization}
h_{\bm{k}} = h_{\bm{k}}^{(0)}+\frac{V_{pd}V_{dp}}{\mu-\varepsilon_{d}}.
\end{equation}

It is confirmed that the hybridization \textcolor{black}{slightly changes the value of $t^{10}$ and $t^{11}$ and introduces new terms} $t^{03}\left(\cos a k_{x}-\cos a k_{y}\right) \tau_{0}\otimes\sigma_{3}$ and $d^{00}_{\bm{k}}\tau_{0}\otimes\sigma_{0}$. The new terms describe the second nearest-neighbor hopping mediated by the $d$-block elements. Interestingly, the direct second nearest-neighbor hopping produces exactly the same functional form due to the glide symmetry.

Our DFT calculation shows that Zr\textit{AQ}-family has stronger hybridization strength than the \textit{AE}Mn\textit{B}$_{2}$-family. Even though there are two less electrons in the Si-net, a strong hybridization modifies the low-energy structure similar to the CaBi(Sb) lattice with the direct second nearest-neighbor hopping. Note that the low-energy band structure is reasonably reproduced with the lowest order of perturbation, since its origin coincides with the second nearest-neighbor hopping. With this point of view, quarter-filled case becomes the effective half-filled \textcolor{black}{system} with smaller band width. As shown in Fig. \hyperref[fig:dispersion]{\ref*{fig:dispersion} \textbf{d}}, perturbative approach seems to be plausible within the energy window including nonsymmorphic degeneracies at X(Y) point.  Since we are dealing with two sites and two orbitals, Hamiltonian can be spanned in the basis of Kronecker products of the Pauli matrices $g_{ij}\equiv \tau_{i}\otimes\sigma_{j}$. Effective Hamiltonian density for the square-net substructure in \textit{P4/nmm} materials reads,

\begin{equation}
\begin{aligned}
h_{\bm{k}}=&t^{10} \cos\frac{a k_{x}}{2}\cos\frac{a k_{y}}{2}g_{10}\\
+&t^{11} \sin\frac{a k_{x}}{2}\sin\frac{a k_{y}}{2}g_{11}\\
+&\frac{\Delta}{4} \left(\cos a k_{x}-\cos a k_{y}\right) g_{03}.
\end{aligned}\end{equation}

Note that we drop the $d^{00}_{\bm{k}}g_{00}$ to ensure the electron-hole symmetry. As mentioned earlier, the second nearest-neighbor hopping also introduces the electron-hole asymmetry ($eh$-asymmetry) from $d^{00}_{\bm{k}}g_{00}$ since there is no anticommuting matrix with the identity. We drop $g_{00}$-related term since the identity matrix $g_{00}$ cannot change the topology of the system and the DFT gives the almost $eh$-symmetric result for ZrSiS and CaMnBi$_{2}$. We will revisit the effect of $eh$-asymmetry potential on the Fermi surface shape together with the effect of SOC. Collaboration of $eh$-asymmetry and SOC may \textcolor{black}{be responsible for} the non-trivial Berry phase in experiment. \cite*{Huang2017,Liu2016} We will discuss this topic further in section \hyperref[sec:berry]{\ref*{sec:berry}}. Without $g_{00}$-related term, $eh$-symmetry exists by the virtue of the chiral-like operator $g_{31}$ that anticommutes with the effective Hamiltonian density.

In terms of $pd$-hybridization \textcolor{black}{parameters}, $\Delta=(2 V_{pd\pi}^{2}-V_{pd\sigma}^{2})/(\mu-\varepsilon_{d})$. On the other hand, $\Delta=4\left(t_{2\sigma}-t_{2\pi}\right)$ in terms of the second nearest-neighbor hopping parameters. In the electronic structure, $\Delta$ is the energy difference between two nonsymmorphic degeneracies at $(\pi/a,0)$ or $(0,\pi/a)$ as shown in Fig. \hyperref[fig:dispersion]{\ref*{fig:dispersion} \textbf{a}} marked by red dots.  Square-net type CaMnBi$_{2}$ also has non-zero $\Delta$ but the value is small. Interestingly, we find that not only $d$-orbitals at 2c site but also $s$- and $p$- orbitals yield the same functional form of the hybridization term up to second order hopping process. It makes sense since one can translate the virtual hopping process to the mediated-second nearest-neighbor hopping. There is suggestion based on DFT calculation that $p$-mediated second nearest-neighbor hopping in BiF lattice that shares the similar low-energy structure\cite*{Luo2015}.

\begin{figure}[tbp]
\centering
\includegraphics[width=1.0\linewidth]{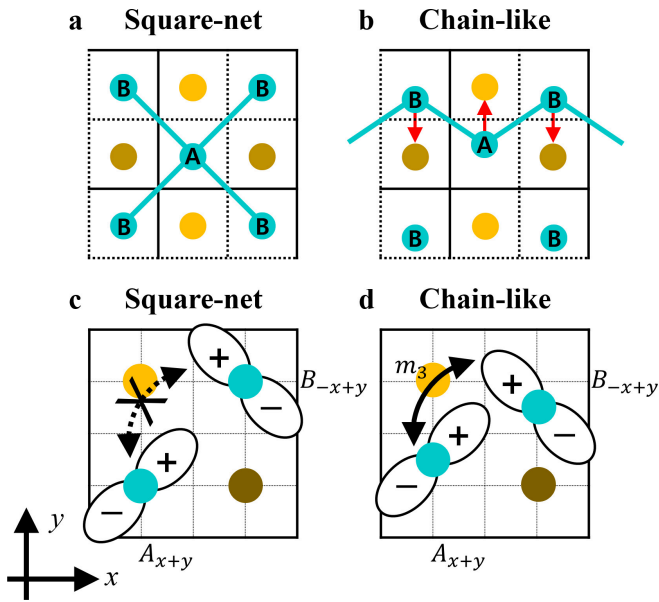}
\caption{(Color online) Schematic representation of \textbf{a} the square-net and \textbf{b} the chain-like structure. UCs are represented by black solid lines. As mentioned in the main text, buckling along $z$-direction is neglected to write the effective Hamiltonian in modified $p_{x,y}$ basis. Due to the lower-symmetry of the chain-like structure, previously forbidden hopping \textbf{c} between $A_{x+y}=(A_{x}+A_{y})/\sqrt{2}$ orbital and $B_{-x+y}=(B_{x}+B_{y})/\sqrt{2}$ orbitals becomes possible as shown in \textbf{d}.} 
\label{fig:variation}
\end{figure}

\subsection{Chain-formation potential}\label{subsec:chain}

Since the materials with chain-like substructure are centrosymmetric \cite*{He2017b,Farhan2014} we force the inversion-symmetric chain-formation in 2-dimensional model as shown in the Fig. \hyperref[fig:variation]{\ref*{fig:variation}}. To restrict the effective Hamiltonian in $p_{x,y}$ subspace, we neglect buckling along $z$-direction. Moreover, we put the effect of orthogonal UC into square UC by changing the hopping strength between inter-chain atoms and intra-chain atoms. Derivation of the chain-formation \textcolor{black}{potential} $(V_{\textrm{chain}})$ shares the same procedure with that of Peierls model in 1-dimensional problem. As mentioned before, we consider the variation along $y$-direction and zigzag-shaped chain is formed along $x$-direction. Difference from the Peierls model is the loss of chiral symmetry after the chain-formation. $V_{\textrm{chain}}$ \textcolor{black}{has} three parameters carried by $g_{20},g_{21},g_{23}$ 
\begin{equation}
\begin{aligned}
V_{\textrm{chain}}
=&\delta t^{10} \cos\frac{a k_{x}}{2}\sin\frac{a k_{y}}{2}g_{20}\\
+&\delta t^{11} \sin\frac{a k_{x}}{2}\cos\frac{a k_{y}}{2}g_{21}\\
+&m_{3} \cos\frac{a k_{x}}{2}\sin\frac{a k_{y}}{2}g_{23}.
\end{aligned}\end{equation}

$\delta t^{10 (11)}$ is the variation of hopping parameter $t^{10 (11)}$ with respect to the first order of real-space variation. $m_{3}$ is introduced as a new hopping mode due to the lower symmetry of the chain-like structure as shown in Fig. \hyperref[fig:hopping]{\ref*{fig:variation} \textbf{c,d}}. Since $g_{23}$ does not anticommute with chiral-like operator $g_{31}$, exact $eh$-symmetry is lost after the chain-formation. If the crystal symmetries are ignored, the topology of the system is simplified into standard $\mathbb{Z}_{2}$ in 2-dimension with only $T^{2}=-1$ symmetry cases\cite*{Kitaev2009,Ryu2010}. To describe the electronic structure evolution under the $V_{\textrm{chain}}$, we defined new parameters from the chain-formation parameters $M_{\pm}\equiv \delta t^{10}\pm m_{3}$. $M_{+(-)}$ lifts the Y-point nonsymmorphic degeneracy below (above) the chemical potential. One last paramter $\delta t^{11}$ mixes the different site $(\tau_{2})$ and orbital $(\sigma_{1})$, and this term lifts the degeneracy along the nodal-line except two points on $\Gamma$-Y. In the later section we will discuss the effect of $V_{\textrm{chain}}$ on (i) the $\mathbb{Z}_{2}$ topology of the system and (ii) the evolution of low-energy band structure and Berry phase on the Fermi surfaces.


\begin{table}[tbp]
\centering
\begin{ruledtabular}
\begin{tabular}{ccccccc}
Coefficient [eV]                         & $t^{10}$ & $t^{11}$ & $\Delta$ & $M_{+}$ & $M_{-}$ & $\delta t^{11}$ \\ \hline
ZrSiS$^{\dagger}$    & $2.00$ & $-4.00$ & $1.40$ & $-$       & $-$     & $-$     \\ 
ZrSnTe$^{\dagger}$ & $1.50$ & $-3.00$ & $1.30$ & $-$       & $-$     & $-$     \\ 
CaMnBi$_{2} $          & $2.40$ & $-4.24$ & $0.16$ & $-$       & $-$     & $-$     \\ 
CaMnSb$_{2}$          & $2.61$ & $-4.60$ & $0.92$ & $0.82$ & $-0.10$ & $0.48$ \\
SrMnSb$_{2} $          & $2.46$ & $-4.28$ & $0.64$ & $0.65$ & $-0.09$ & $0.37$ \\
\end{tabular}
\end{ruledtabular}
\caption{Tight binding parameters extracted from the PBE-GGA+mBJ calculations for selected materials. $^{\dagger}$ Paramters are valid near the chemical potential due to the strong $pd$-hybridization.}
\label{table:parameters}
\end{table}

\section{Topological phase diagram}\label{sec:phase}

\begin{figure}[bp]
\centering
\includegraphics[width=1.0\linewidth]{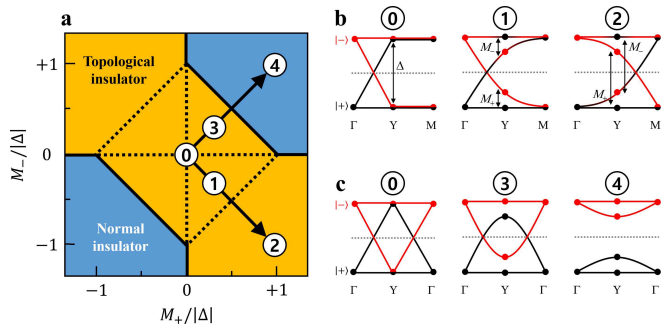}
\caption{(Color online) \textbf{a} Phase diagram based on parity eigenvalues. There are two distinct regions of topological insulator (yellow, $\mathbb{Z}_{2}=1$) and normal insulator (blue, $\mathbb{Z}_{2}=0$). Black solid lines separate the topologically distinct regions and black dashed lines separate regions with different band-connectivity. Circled numbers are given to different band-connectivity as seen in \textbf{b} and \textbf{c}. Note that the circled numbers denote only the weak- and strong-chain regions not the specific point on the phase diagram. In \textbf{b} and \textbf{c}, black (red) solid dots present states with parity eigenvalue of $+1(-1)$, and gap parameters are presented. Schematic band evolution along topologically non-trivial regime is given in \textbf{b}. Topological transition is schematically represented in \textbf{c}, there is parity exchange at M point. Note that the SOC is neglected.} 
\label{fig:phase}
\end{figure}


We numerically calculated $\mathbb{Z}_{2}$ invariant of the system. To do this, we used the Hamiltonian written in UC and the parity operator given by $-g_{10}$ that maps $A(B)_{x,y}$ to $-B(A)_{x,y}$. From the calculated parity eigenvalues at \textcolor{black}{time-reversal invariant momenta (TRIM)}, we build the topological phase diagram with respect to the chain-formation parameters $M_{\pm}$. Specific symmetry of the topological phase diagram is due to the ambiguity of the atomic-variational direction in Fig. \hyperref[fig:phase]{\ref*{fig:phase} \textbf{a}}. Circled numbers only assign to the regions with different band-connectivity not the specific value on the phase diagram. For example, \circled{2} represents two disconnected regions where $M_{+}$ and $M_{-}$ have opposite sign and $\left|M_{+}\right|+\left|M_{-}\right|> \left|\Delta\right|$ \textcolor{black}{and \circled{4} represents regions where $M_{+}$ and $M_{-}$ have the same sign and $\left|M_{+}\right|+\left|M_{-}\right|> \left|\Delta\right|$.}

Without $V_{\textrm{chain}}$ shown in Fig. \hyperref[fig:phase]{\ref*{fig:phase} \textbf{b-c}}, the parity exchange occurs from $\Gamma$ to other three TRIM to make $\mathbb{Z}_{2}=1$. This band crossing forms nodal-line semimetal at the center \circled{0}. In this point of view, topological phase transition may occur when chain-formation parameters $M_{\pm}$ overcomes the hybridization strength $\Delta$ to exchange the parity eigenstates. In other words, $d$-block ions at 2c site not only generate specific symmetries of the system but also protect the non-trivial topology of the system. Moreover, real material favors the TI phase since the presence of $m_{3}$ ensures the gap opening of the opposite sign at $Y$ point. Since the gap along the nodal-line is readily available through the SOC, topological surface-state signals in various material are promising.  We will briefly discuss the availability of the global gap from the Kane-Mele type SOC in the next section.

We sketch the evolution of the band connectivity along $\Gamma-\mathrm{Y}-\mathrm{M}\ (\Gamma)$ in Fig. \hyperref[fig:phase]{\ref*{fig:phase} \textbf{b} (\textbf{c})}. In the case of square-net, where the set of band crossing points form a nodal-line, the parity eigenstates are exchanged from $\Gamma$ to other three TRIM. Note that real materials favor the regions \circled{0}, \circled{1}, and \circled{2}. Square-net materials like ZrSiQ (Q=S,Se,Te), ZrSnTe, and CaMnBi$_{2}$ are located at the center \circled{0} of the phase diagram. In the case of \circled{1} or \circled{3}, where the hybridization gap at Y point $(\Delta)$ is bigger than the sum of the chain-formation parameters \textcolor{black}{$\left|M_{+}\right|+\left|M_{-}\right|< \left|\Delta\right|$}, the parity exchange does not occur and $\mathbb{Z}_{2}$ remains the same with the square-net. We will call these as weak-chain systems. Apparently, there seems no difference between the same- and opposite-sign cases. There exists a single crossing along $\Gamma$-Y and the other part is gapped due to non-zero $\delta t^{11}$. We will visit the difference between case \protect\circled{1} and case \protect\circled{3} briefly at section \hyperref[sec:berry]{\ref*{sec:berry}}.

In the strong-chain systems of \protect\circled{2} and \protect\circled{4}, the chain formation parameters overcome the hybridization gap \textcolor{black}{$\left|M_{+}\right|+\left|M_{-}\right|> \left|\Delta\right|$}. In region \protect\circled{4}, the chain formation parameters have the same sign and they overcome the hybridization gap, so the topological transition occurs and the material is fully gapped by the chain-formation parameters. Besides, there exists a stable TI phase \protect\circled{2} protected by the chain-formation. \textcolor{black}{There is topological critical point between the two topologically distinct regions. Collection of the topological critical points in the phase space is represented as black solid lines in Fig. \hyperref[fig:phase]{\ref*{fig:phase} \textbf{a}}. On these topological critical points $\left|M_{+}\right|+\left|M_{-}\right|= \left|\Delta\right|$, linear band touching occurs at Y point and $\mathbb{Z}_{2}$ invariant is ill-defined. In the real materials, TI phase is strongly favorable since the low-symmetry of the chain-like structure always \textcolor{black}{carries} non-zero $m_{3}$.} This is the case for CaMnSb$_{2}$ where one of the mass gap is inverted so the parity eigenstate cannot be exchanged by increasing the chain-formation parameters. This state is characterized by the band crossing along Brillouin zone boundary Y-M. Thus, it is possible to observe the surface states from the materials containing square-net or chain-like substructure.

\section{Berry hot spot}\label{sec:berry}

In the previous section, we studied the fact that band-connectivity between the TRIM determines the topology of the system since it is centrosymmetric. Reversely, one can classify the topology of the system by investigating the shape of the Fermi surfaces since the shape changes according to the band connectivity. In particular, one can see that the location of the Fermi surface changes significantly by adjusting the chain formation parameters. In this section, we classify the shape of the Fermi surface and discuss its role as an origin of the Berry phase in SdH experiments. We will call the Fermi surfaces that enclose the \emph{large} Berry curvature as Berry hot spots.

\subsection{Square-net system}

First, we numerically calculate the Berry phase using $h_{\bm{k}}$ along the Fermi surface\cite*{Lee2016} derived from the $eh$-symmetric nodal-line semimetal by adjusting the Fermi energy ($E_{\textrm{F}}$). \textcolor{black}{We set $E_{\textrm{F}}$ between the nodal-line degeneracy and the nonsymmorphic degeneracies to make the closed Fermi surfaces.} We find that the Berry phase is $2\pi$ on the Fermi surface of the $eh$-symmetric system. Even though the Fermi surface has the non-trivial topology in terms of the pseudospin winding number, the Berry phase is trivial. Here, the pseudospin winding number is defined according to Ref.[\onlinecite*{Lee2016}]. 


Then the non-trivial Berry phase should be the result of the $eh$-asymmetry and the SOC can change the topology of the Fermi surface by interweaving and cutting the band structure at $E_{\textrm{F}}$. \cite*{Huang2017,Liu2016}. \textcolor{black}{To explore the relativistic effect on the low-energy state, we calculate the matrix elements of SOC in $p_{x,y,z}$ basis by using the relation between the spherical harmonics $Y_{l}^{m}$ and real harmonics $p_{x}=(Y_{1}^{1}+Y_{1}^{-1})/\sqrt{2}$, $p_{x}=i(Y_{1}^{1}-Y_{1}^{-1})/\sqrt{2}$, and $p_{z}=Y_{1}^{0}$. We can write the SOC matrix for $p$-orbitals SOC$_{p}$ in terms of the following basis,
\begin{equation*}
\left|p_{x,\uparrow}\right>,\ \left|p_{x,\downarrow}\right>,\ 
\left|p_{y,\uparrow}\right>,\ \left|p_{y,\downarrow}\right>,\ 
\left|p_{z,\uparrow}\right>,\ \left|p_{z,\downarrow}\right>
\end{equation*}
where $\uparrow(\downarrow)$ indicate the spin. With this basis, $SOC_{p}$ becomes
\begin{equation}
\textrm{SOC}_{p}=\lambda_{p}\ 
\left(\begin{array}{cccccc} 
0 & 0 & i & 0 & 0 & 1 \\ 
0 & 0 & 0&-i &-1& 0 \\ 
-i& 0 & 0 & 0 & 0 & i \\ 
0 & i & 0 & 0 & i & 0 \\ 
0 &-1 & 0 &-i & 0 & 0 \\ 
1 & 0 &-i & 0 & 0 & 0 \\ 
\end{array}\right).
\end{equation}
Since we restrict the effective Hamiltonian in $p_{x,y}$ basis, the $p_{x,y}$ part of the full atomic SOC is projected out as $\textrm{SOC}_{0}=\lambda_{0}\ \tau_{0}\otimes\sigma_{2}\otimes s_{3}$ where Pauli matrices for spin $s_{i}$ are introduced. In this way, we completely lose the interaction from the $p_{z}$ state.} In order to include the effect from $p_{z}$ state, Kane-Mele type second order SOC \cite*{Kane2005a} is considered in addition to the atomic SOC projected on $p_{x,y}$ orbitals. Explicit form of the Kane-Mele type SOC can be obtained from the following hopping path containing Zr- or Ca-$d$ and $p_{z}$ orbitals. Note that $\alpha$ and $\beta$ are the indices for $A(B)_{x,y}$ orbitals and $\gamma$ is the index for $d$-orbitals.

\begin{equation}
V_{pd}^{\ga\gamma}d_{\gamma s}d_{\gamma s}^{\dagger}V_{dp}^{\gamma z}p_{zs}p_{zs}^{\dagger}\textrm{SOC}_{p}^{z\gb ss'}+h.c.
\end{equation}

We then select the mirror-symmetric SOC ($\textrm{SOC}_{\textrm{KM}}$) potential that agrees with the geometric interpretation of the SOC\cite*{Min2006, Liu2011b}. This SOC does not change the form under the chain-formation since it connects the second nearest-neighbor.

\begin{equation}
\textrm{SOC}_{\textrm{KM}}=\lambda_{\textrm{KM}}\ g_{31}\otimes\left(\sin a k_{x} s_{1}-\sin a k_{y} s_{2}\right)\\
\end{equation}

\begin{figure}[htbp]
	\centering
	\includegraphics[width=1.0\linewidth]{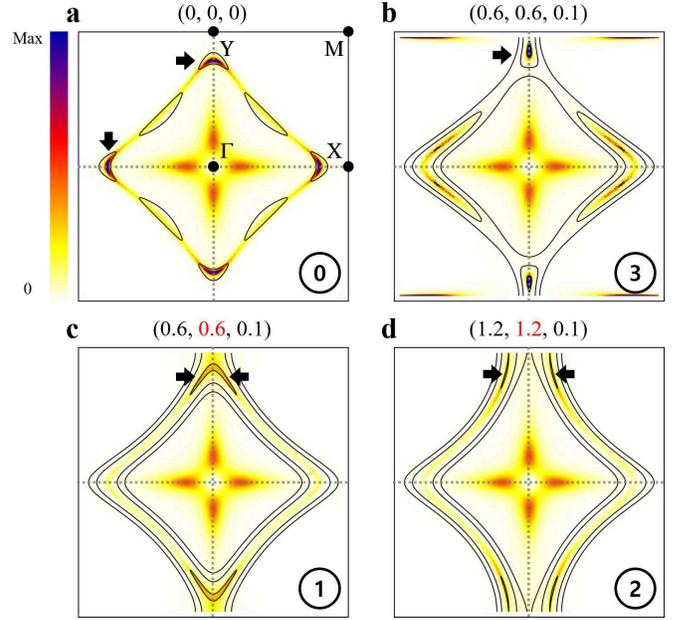}
	\caption{\textbf{a-d} Numerically calculated absolute value of the spin-dependent Berry curvature difference $\left|\Omega_{s}-\Omega_{\bar{s}}\right|$ of low-energy valence band in topologically non-trivial regime. $(t^{10},t^{11},\Delta,\lambda_{\textrm{KM}}, \lambda_{\textrm{0}})=(2.0, -4.0, 0.7,0.1,0.1)$ in eV is used for all the figures. Chain-formation parameters $(M_{+},M_{-}, \delta t^{11})$ in eV are given above in each figure. Black and red texts indicate the positive and negative values, respectively. Black arrows indicate the positions of $\left|\Omega_{s}-\Omega_{\bar{s}}\right|$ maxima, Berry hot spots. Colormap is scaled with respect to square-net system shown in \textbf{a}. Black solid lines are isoenergy contours separated by 0.1 eV for the low-energy valence band. \textcolor{black}{These black solid lines can be interpreted roughly as the possible shape of the Fermi surfaces in the real materials. High-symmetry points or TRIM are indicated as a black dots in \textbf{a}.} Note that we restricted the crystal momentum value to avoid Berry curvature singularities from nonsymmorphic degeneracies at $X$ and $M$. The numbers in circles are used as in Fig. \hyperref[fig:phase]{\ref*{fig:phase}}.}
	\label{fig:berrycurvature}
\end{figure}

\textcolor{black}{After the inclusion of all the effect from the chain-formation and SOC, total Hamiltonian density $H_{\bm{k}}$ becomes
\begin{equation}\begin{aligned}
H_{\bm{k}}=\left(h_{\bm{k}}+V_{\textrm{chain}}\right)\otimes s_{0}+\textrm{SOC}_{\textrm{KM}}+\textrm{SOC}_{0}.
\end{aligned}\end{equation}}
Since the Berry phase $(\gamma)$ of the Fermi surface is the integration of the Berry curvature $(\Omega)$ surrounded by the Fermi surface, it severly depends on $E_{\textrm{F}}$ and $eh$-asymmetry. We present only the absolute value of the spin-dependent Berry curvature difference $\left|\Omega_{s}-\Omega_{\bar{s}}\right|$ of low-energy valence band in Fig.\hyperref[fig:berrycurvature]{\ref*{fig:berrycurvature}}.

For the perfect square-net systems \protect\circled{0}, the nodal-line is fully gapped and inclusion of $eh$-asymmetry \textcolor{black}{produces} eight Fermi surfaces. Within these Fermi surfaces, hot spots for Berry phase are the Fermi surfaces lying on $\Gamma$-X and $\Gamma$-Y, as shown in Fig.\hyperref[fig:berrycurvature]{\ref*{fig:berrycurvature} \textbf{a}}. \textcolor{black}{We predict that these four hot spots are the main source of the Berry phase in Zr$AQ$-family\cite*{Ali2016,Hu2016}}. Note that the curvatures near the $\Gamma$ points are coming from the high-energy state and does not affect the transport property of the system.


\subsection{Chain-like system}

We follow the evolution path presented in Fig. \hyperref[fig:phase]{\ref*{fig:phase} \textbf{b}}. As explained before, weak chain-formation opens gap associated with $\delta t^{11}$ along the nodal-line. However, there remain two points along $\Gamma$-Y within the Brillouin zone where the gap becomes 0 \textcolor{black}{as shown in case \protect\circled{1} of Fig. \hyperref[fig:phase]{\ref*{fig:phase} \textbf{b}}}. If we turn on the remaining chain-formation parameters $M_{\pm}$, the crossing point itself also moves toward Y point. For the low-energy valence band, Berry curvature is transferred  from the gap located on $\Gamma$-X to the SOC-gap located on $\Gamma$-Y as shown in Fig. \hyperref[fig:berrycurvature]{\ref*{fig:berrycurvature} \textbf{c}} compared with the square-net in Fig.\hyperref[fig:berrycurvature]{\ref*{fig:berrycurvature} \textbf{a}}. Thus, two banana-shaped Fermi surfaces centered on $\Gamma$-Y \textcolor{black}{that derived from the band crossing points of case \protect\circled{1}} will be the main source of the experimental Berry phase signal.  Up to our knowledge, there is no known material in this weak chain class. But PBE-GGA+mBJ calculation suggests that CaMnSb$_{2}$ located on the boundary between weak chain \protect\circled{1} and strong chain \protect\circled{2} as shown in Table \hyperref[table:parameters]{\ref{table:parameters}}.

During the evolution, the crossing points move toward Y point and merge together at Y point when \textcolor{black}{$\left|M_{+}\right|+\left|M_{-}\right|= \left|\Delta\right|$.} At this point \textcolor{black}{between case \protect\circled{1} and \protect\circled{2}}, one loose the Berry phase signal since the phase approaches to $2\pi$ similar to the square-net case.  If the chain-formation parameters overcome the hybridization $\Delta$, the separation occurs along $(\xi,\pi/a)$ as shown in case  \protect\circled{2} of Fig. \hyperref[fig:phase]{\ref*{fig:phase} \textbf{b}}. The Berry hot spots experience the same movement since the hot spots are located at the SOC gap between the band crossing. Hence, the hot spots of strong-chain system \protect\circled{2} also located near the zone boundary as shown in Fig. \hyperref[fig:berrycurvature]{\ref*{fig:berrycurvature} \textbf{d}}. As explained before, this evolution process does not alter the parity product at Y point and the material remains topologically non-trivial. In the PBE-GGA+mBJ calculation without SOC, crossing point along $(\xi,\pi/a)$ can be observed in SrMnSb$_2$. Schematic band connectivity of SrMnSb$_{2}$ is presented in Fig. \hyperref[fig:phase]{\ref*{fig:phase} \textbf{b}}.


Finally, we briefly discuss the unfavorable transition path presented in Fig. \hyperref[fig:phase]{\ref*{fig:phase} \textbf{c}}. Like the weak-chain TI \protect\circled{1}, there is no strong-chain NI \protect\circled{4} up to our current knowledge. Band crossing along $\Gamma$-Y seems similar as shown in Fig. \hyperref[fig:phase]{\ref*{fig:phase} \textbf{b-c}}, detailed electronic structures are different from each other as shown in the isoenergy contour in \hyperref[fig:berrycurvature]{\ref*{fig:berrycurvature} \textbf{b-c}}. As discussed before, weak chain-formation opens gap associated with $m_{1}$ along nodal-line. However, there remain two points along $\Gamma$-Y within the Brillouin zone where the gap is zero. These linear crossings \textcolor{black}{provide} strong Berry hot spots, i.e. the concentrated Berry curvature, as shown in \hyperref[fig:berrycurvature]{\ref*{fig:berrycurvature} \textbf{b}}.

\section{Discussion}\label{sec:discussion}
\textcolor{black}{
There is serious material dependency in the $eh$-asymmetry\cite*{Li2016}. The contribution carried by the identity $\left(g_{00}\right)$ trivially gives $eh$-asymmetry since it does not anticommute with any other matrices. Beyond that, the inclusion of the third nearest-neighbor hopping processes \textcolor{black}{generates} the non-trivial $eh$-asymmetry \cite*{Inho} by adding single $g_{01}$ term. Together with this term, there is no anticommuting matrix $g_{ij}$ for Hamiltonian containing the first, second, and third nearest-neighbor hopping processes without distortion. It seems that a precise description including $eh$-asymmetry needs the higher order hopping processes. Since the $eh$-asymmetry and SOC interweave and cut the nodal-line into Fermi surfaces that confine the non-trivial Berry curvature, the material dependency of the $eh$-asymmetry should be resolved further to precisely calculate the transport property of such materials.}

\section{Conclusion}\label{sec:conclusion}

Based on the DFT electronic band structures, we construct TB model for square-net \textit{P4/nmm} materials and chain-like \textit{Pnma} materials. From the TB analysis, hybridization with ions at 2c site is not material specific since $s$-, $p$-, and $d$-orbitals at 2c give the same hybridization potential within the second order perturbation theory. Square-net Zr$AQ$-family and $AE$Mn$B$$_{2}$-family share the same effective Hamiltonian near the chemical potential, since the hybridization-mediated hopping has the same functional form with a direct second nearest-neighbor hopping due to the glide symmetry.

Topological phase transition occurs when the chain-formation parameters overcome the hybridization gap between the nonsymmorphic degeneracies at $(\pm\pi/a,0)$ or $(0,\pm\pi/a)$. In this point of view, the topology of the system is protected by the hybridization itself. \textcolor{black}{Moreover, newly introduced mass $m_{3}$ due to low-symmetry of the chain-like structure favors the TI phase.} Hence, every member of Zr$AQ$- and $AE$Mn$B$$_{2}$-family is all potential TI as long as \textcolor{black}{the global SOC gap is present.} Depending on $E_{\textrm{F}}$, Fermi surfaces with non-zero Berry phase can be observed. These Berry hot spots are very material specific and have correspondence with the topology of the system since the low-energy band structure and topology of the system vary with the chain-formation parameters. 
\\
\begin{acknowledgments}
This research was supported by Basic Science Research Program through the National Research Foundation of Korea(NRF) funded by the Ministry of Education (NRF-2017R1D1A1B03032069).
\end{acknowledgments}




\bibliographystyle{apsrev4-1}

\begin{thebibliography}{41}%
\makeatletter
\providecommand \@ifxundefined [1]{%
 \@ifx{#1\undefined}
}%
\providecommand \@ifnum [1]{%
 \ifnum #1\expandafter \@firstoftwo
 \else \expandafter \@secondoftwo
 \fi
}%
\providecommand \@ifx [1]{%
 \ifx #1\expandafter \@firstoftwo
 \else \expandafter \@secondoftwo
 \fi
}%
\providecommand \natexlab [1]{#1}%
\providecommand \enquote  [1]{``#1''}%
\providecommand \bibnamefont  [1]{#1}%
\providecommand \bibfnamefont [1]{#1}%
\providecommand \citenamefont [1]{#1}%
\providecommand \href@noop [0]{\@secondoftwo}%
\providecommand \href [0]{\begingroup \@sanitize@url \@href}%
\providecommand \@href[1]{\@@startlink{#1}\@@href}%
\providecommand \@@href[1]{\endgroup#1\@@endlink}%
\providecommand \@sanitize@url [0]{\catcode `\\12\catcode `\$12\catcode
  `\&12\catcode `\#12\catcode `\^12\catcode `\_12\catcode `\%12\relax}%
\providecommand \@@startlink[1]{}%
\providecommand \@@endlink[0]{}%
\providecommand \url  [0]{\begingroup\@sanitize@url \@url }%
\providecommand \@url [1]{\endgroup\@href {#1}{\urlprefix }}%
\providecommand \urlprefix  [0]{URL }%
\providecommand \Eprint [0]{\href }%
\providecommand \doibase [0]{http://dx.doi.org/}%
\providecommand \selectlanguage [0]{\@gobble}%
\providecommand \bibinfo  [0]{\@secondoftwo}%
\providecommand \bibfield  [0]{\@secondoftwo}%
\providecommand \translation [1]{[#1]}%
\providecommand \BibitemOpen [0]{}%
\providecommand \bibitemStop [0]{}%
\providecommand \bibitemNoStop [0]{.\EOS\space}%
\providecommand \EOS [0]{\spacefactor3000\relax}%
\providecommand \BibitemShut  [1]{\csname bibitem#1\endcsname}%
\let\auto@bib@innerbib\@empty
\bibitem [{\citenamefont {Young}\ and\ \citenamefont {Kane}(2015)}]{Young2015}%
  \BibitemOpen
  \bibfield  {author} {\bibinfo {author} {\bibfnamefont {S.~M.} \bibnamefont
  {Young}}\ and\ \bibinfo {author} {\bibfnamefont {C.~L.} \bibnamefont
  {Kane}},\ }\href {\doibase 10.1103/PhysRevLett.115.126803} {\bibfield
  {journal} {\bibinfo  {journal} {Phys. Rev. Lett.}\ }\textbf {\bibinfo
  {volume} {115}},\ \bibinfo {pages} {126803} (\bibinfo {year}
  {2015})}\BibitemShut {NoStop}%
\bibitem [{\citenamefont {Yang}\ \emph {et~al.}(2017)\citenamefont {Yang},
  \citenamefont {Bojesen}, \citenamefont {Morimoto}, and\ \citenamefont
  {Furusaki}}]{Yang2017}%
  \BibitemOpen
  \bibfield  {author} {\bibinfo {author} {\bibfnamefont {B.-J.} \bibnamefont
  {Yang}}, \bibinfo {author} {\bibfnamefont {T.~A.} \bibnamefont {Bojesen}},
  \bibinfo {author} {\bibfnamefont {T.} \bibnamefont {Morimoto}}, and\
  \bibinfo {author} {\bibfnamefont {A.} \bibnamefont {Furusaki}},\ }\href
  {\doibase 10.1103/PhysRevB.95.075135} {\bibfield  {journal} {\bibinfo
  {journal} {Phys. Rev. B}\ }\textbf {\bibinfo {volume} {95}},\ \bibinfo
  {pages} {075135} (\bibinfo {year} {2017})}\BibitemShut {NoStop}%
\bibitem [{\citenamefont {Park}\ and\ \citenamefont {Yang}(2017)}]{Park2017}%
  \BibitemOpen
  \bibfield  {author} {\bibinfo {author} {\bibfnamefont {S.} \bibnamefont
  {Park}}\ and\ \bibinfo {author} {\bibfnamefont {B.-J.} \bibnamefont
  {Yang}},\ }\href {\doibase 10.1103/PhysRevB.96.125127} {\bibfield  {journal}
  {\bibinfo  {journal} {Phys. Rev. B}\ }\textbf {\bibinfo {volume} {96}},\
  \bibinfo {pages} {125127} (\bibinfo {year} {2017})}\BibitemShut {NoStop}%
\bibitem [{\citenamefont {Wang}(2017)}]{Wang2017}%
  \BibitemOpen
  \bibfield  {author} {\bibinfo {author} {\bibfnamefont {J.} \bibnamefont
  {Wang}},\ }\href {\doibase 10.1103/PhysRevB.95.115138} {\bibfield  {journal}
  {\bibinfo  {journal} {Phys. Rev. B}\ }\textbf {\bibinfo {volume} {95}},\
  \bibinfo {pages} {115138} (\bibinfo {year} {2017})}\BibitemShut {NoStop}%
\bibitem [{\citenamefont {Neupane}\ \emph {et~al.}(2016)\citenamefont
  {Neupane}, \citenamefont {Belopolski}, \citenamefont {Hosen}, \citenamefont
  {Sanchez}, \citenamefont {Sankar}, \citenamefont {Szlawska}, \citenamefont
  {Xu}, \citenamefont {Dimitri}, \citenamefont {Dhakal}, \citenamefont
  {Maldonado}, \citenamefont {Oppeneer}, \citenamefont {Kaczorowski},
  \citenamefont {Chou}, \citenamefont {Hasan}, and\ \citenamefont
  {Durakiewicz}}]{Neupane2016}%
  \BibitemOpen
  \bibfield  {author} {\bibinfo {author} {\bibfnamefont {M.} \bibnamefont
  {Neupane}}, \bibinfo {author} {\bibfnamefont {I.} \bibnamefont {Belopolski}},
  \bibinfo {author} {\bibfnamefont {M.~M.} \bibnamefont {Hosen}}, \bibinfo
  {author} {\bibfnamefont {D.~S.} \bibnamefont {Sanchez}}, \bibinfo {author}
  {\bibfnamefont {R.} \bibnamefont {Sankar}}, \bibinfo {author} {\bibfnamefont
  {M.} \bibnamefont {Szlawska}}, \bibinfo {author} {\bibfnamefont {S.-Y.}\
  \bibnamefont {Xu}}, \bibinfo {author} {\bibfnamefont {K.} \bibnamefont
  {Dimitri}}, \bibinfo {author} {\bibfnamefont {N.} \bibnamefont {Dhakal}},
  \bibinfo {author} {\bibfnamefont {P.} \bibnamefont {Maldonado}}, \bibinfo
  {author} {\bibfnamefont {P.~M.} \bibnamefont {Oppeneer}}, \bibinfo {author}
  {\bibfnamefont {D.} \bibnamefont {Kaczorowski}}, \bibinfo {author}
  {\bibfnamefont {F.} \bibnamefont {Chou}}, \bibinfo {author} {\bibfnamefont
  {M.~Z.} \bibnamefont {Hasan}}, and\ \bibinfo {author} {\bibfnamefont
  {T.} \bibnamefont {Durakiewicz}},\ }\href {\doibase
  10.1103/PhysRevB.93.201104} {\bibfield  {journal} {\bibinfo  {journal} {Phys.
  Rev. B}\ }\textbf {\bibinfo {volume} {93}},\ \bibinfo {pages} {201104}
  (\bibinfo {year} {2016})}\BibitemShut {NoStop}%
\bibitem [{\citenamefont {Hosen}\ \emph {et~al.}(2017)\citenamefont {Hosen},
  \citenamefont {Dimitri}, \citenamefont {Belopolski}, \citenamefont
  {Maldonado}, \citenamefont {Sankar}, \citenamefont {Dhakal}, \citenamefont
  {Dhakal}, \citenamefont {Cole}, \citenamefont {Oppeneer}, \citenamefont
  {Kaczorowski}, \citenamefont {Chou}, \citenamefont {Hasan}, \citenamefont
  {Durakiewicz}, and\ \citenamefont {Neupane}}]{Hosen2017}%
  \BibitemOpen
  \bibfield  {author} {\bibinfo {author} {\bibfnamefont {M.~M.} \bibnamefont
  {Hosen}}, \bibinfo {author} {\bibfnamefont {K.} \bibnamefont {Dimitri}},
  \bibinfo {author} {\bibfnamefont {I.} \bibnamefont {Belopolski}}, \bibinfo
  {author} {\bibfnamefont {P.} \bibnamefont {Maldonado}}, \bibinfo {author}
  {\bibfnamefont {R.} \bibnamefont {Sankar}}, \bibinfo {author} {\bibfnamefont
  {N.} \bibnamefont {Dhakal}}, \bibinfo {author} {\bibfnamefont
  {G.} \bibnamefont {Dhakal}}, \bibinfo {author} {\bibfnamefont
  {T.} \bibnamefont {Cole}}, \bibinfo {author} {\bibfnamefont {P.~M.}\
  \bibnamefont {Oppeneer}}, \bibinfo {author} {\bibfnamefont {D.} \bibnamefont
  {Kaczorowski}}, \bibinfo {author} {\bibfnamefont {F.} \bibnamefont {Chou}},
  \bibinfo {author} {\bibfnamefont {M.~Z.} \bibnamefont {Hasan}}, \bibinfo
  {author} {\bibfnamefont {T.} \bibnamefont {Durakiewicz}}, and\ \bibinfo
  {author} {\bibfnamefont {M.} \bibnamefont {Neupane}},\ }\href {\doibase
  10.1103/PhysRevB.95.161101} {\bibfield  {journal} {\bibinfo  {journal} {Phys.
  Rev. B}\ }\textbf {\bibinfo {volume} {95}},\ \bibinfo {pages} {161101}
  (\bibinfo {year} {2017})}\BibitemShut {NoStop}%
\bibitem [{\citenamefont {Hu}\ \emph {et~al.}(2016)\citenamefont {Hu},
  \citenamefont {Tang}, \citenamefont {Liu}, \citenamefont {Liu}, \citenamefont
  {Zhu}, \citenamefont {Graf}, \citenamefont {Myhro}, \citenamefont {Tran},
  \citenamefont {Lau}, \citenamefont {Wei}, and\ \citenamefont
  {Mao}}]{Hu2016}%
  \BibitemOpen
  \bibfield  {author} {\bibinfo {author} {\bibfnamefont {J.} \bibnamefont
  {Hu}}, \bibinfo {author} {\bibfnamefont {Z.} \bibnamefont {Tang}}, \bibinfo
  {author} {\bibfnamefont {J.} \bibnamefont {Liu}}, \bibinfo {author}
  {\bibfnamefont {X.} \bibnamefont {Liu}}, \bibinfo {author} {\bibfnamefont
  {Y.} \bibnamefont {Zhu}}, \bibinfo {author} {\bibfnamefont {D.} \bibnamefont
  {Graf}}, \bibinfo {author} {\bibfnamefont {K.} \bibnamefont {Myhro}},
  \bibinfo {author} {\bibfnamefont {S.} \bibnamefont {Tran}}, \bibinfo {author}
  {\bibfnamefont {C.~N.} \bibnamefont {Lau}}, \bibinfo {author} {\bibfnamefont
  {J.} \bibnamefont {Wei}}, and\ \bibinfo {author} {\bibfnamefont
  {Z.} \bibnamefont {Mao}},\ }\href {\doibase 10.1103/PhysRevLett.117.016602}
  {\bibfield  {journal} {\bibinfo  {journal} {Phys. Rev. Lett.}\ }\textbf
  {\bibinfo {volume} {117}},\ \bibinfo {pages} {016602} (\bibinfo {year}
  {2016})}\BibitemShut {NoStop}%
\bibitem [{\citenamefont {Ali}\ \emph {et~al.}(2016)\citenamefont {Ali},
  \citenamefont {Schoop}, \citenamefont {Garg}, \citenamefont {Lippmann},
  \citenamefont {Lara}, \citenamefont {Lotsch}, and\ \citenamefont
  {Parkin}}]{Ali2016}%
  \BibitemOpen
  \bibfield  {author} {\bibinfo {author} {\bibfnamefont {M.~N.} \bibnamefont
  {Ali}}, \bibinfo {author} {\bibfnamefont {L.~M.} \bibnamefont {Schoop}},
  \bibinfo {author} {\bibfnamefont {C.} \bibnamefont {Garg}}, \bibinfo {author}
  {\bibfnamefont {J.~M.} \bibnamefont {Lippmann}}, \bibinfo {author}
  {\bibfnamefont {E.} \bibnamefont {Lara}}, \bibinfo {author} {\bibfnamefont
  {B.} \bibnamefont {Lotsch}}, and\ \bibinfo {author} {\bibfnamefont
  {S.~S.~P.} \bibnamefont {Parkin}},\ }\href {\doibase 10.1126/sciadv.1601742}
  {\bibfield  {journal} {\bibinfo  {journal} {Sci. Adv.}\ }\textbf {\bibinfo
  {volume} {2}},\ \bibinfo {pages} {e1601742} (\bibinfo {year}
  {2016})}\BibitemShut {NoStop}%
\bibitem [{\citenamefont {Schoop}\ \emph {et~al.}(2016)\citenamefont {Schoop},
  \citenamefont {Ali}, \citenamefont {Stra{\ss}er}, \citenamefont {Topp},
  \citenamefont {Varykhalov}, \citenamefont {Marchenko}, \citenamefont
  {Duppel}, \citenamefont {Parkin}, \citenamefont {Lotsch}, and\ \citenamefont
  {Ast}}]{Schoop2015}%
  \BibitemOpen
  \bibfield  {author} {\bibinfo {author} {\bibfnamefont {L.~M.} \bibnamefont
  {Schoop}}, \bibinfo {author} {\bibfnamefont {M.~N.} \bibnamefont {Ali}},
  \bibinfo {author} {\bibfnamefont {C.} \bibnamefont {Stra{\ss}er}}, \bibinfo
  {author} {\bibfnamefont {A.} \bibnamefont {Topp}}, \bibinfo {author}
  {\bibfnamefont {A.} \bibnamefont {Varykhalov}}, \bibinfo {author}
  {\bibfnamefont {D.} \bibnamefont {Marchenko}}, \bibinfo {author}
  {\bibfnamefont {V.} \bibnamefont {Duppel}}, \bibinfo {author} {\bibfnamefont
  {S.~S.~P.} \bibnamefont {Parkin}}, \bibinfo {author} {\bibfnamefont {B.~V.}\
  \bibnamefont {Lotsch}}, and\ \bibinfo {author} {\bibfnamefont {C.~R.}\
  \bibnamefont {Ast}},\ }\href {\doibase 10.1038/ncomms11696} {\bibfield
  {journal} {\bibinfo  {journal} {Nat. Commun.}\ }\textbf {\bibinfo {volume}
  {7}},\ \bibinfo {pages} {11696} (\bibinfo {year} {2016})}\BibitemShut
  {NoStop}%
\bibitem [{\citenamefont {Topp}\ \emph {et~al.}(2016)\citenamefont {Topp},
  \citenamefont {Lippmann}, \citenamefont {Varykhalov}, \citenamefont {Duppel},
  \citenamefont {Lotsch}, \citenamefont {Ast}, and\ \citenamefont
  {Schoop}}]{Topp2016}%
  \BibitemOpen
  \bibfield  {author} {\bibinfo {author} {\bibfnamefont {A.} \bibnamefont
  {Topp}}, \bibinfo {author} {\bibfnamefont {J.~M.} \bibnamefont {Lippmann}},
  \bibinfo {author} {\bibfnamefont {A.} \bibnamefont {Varykhalov}}, \bibinfo
  {author} {\bibfnamefont {V.} \bibnamefont {Duppel}}, \bibinfo {author}
  {\bibfnamefont {B.~V.} \bibnamefont {Lotsch}}, \bibinfo {author}
  {\bibfnamefont {C.~R.} \bibnamefont {Ast}}, and\ \bibinfo {author}
  {\bibfnamefont {L.~M.} \bibnamefont {Schoop}},\ }\href
  {http://stacks.iop.org/1367-2630/18/i=12/a=125014} {\bibfield  {journal}
  {\bibinfo  {journal} {New J. Phys.}\ }\textbf {\bibinfo {volume} {18}},\
  \bibinfo {pages} {125014} (\bibinfo {year} {2016})}\BibitemShut {NoStop}%
\bibitem [{\citenamefont {Topp}\ \emph {et~al.}(2017)\citenamefont {Topp},
  \citenamefont {Queiroz}, \citenamefont {Gr\"uneis}, \citenamefont
  {M\"uchler}, \citenamefont {Rost}, \citenamefont {Varykhalov}, \citenamefont
  {Marchenko}, \citenamefont {Krivenkov}, \citenamefont {Rodolakis},
  \citenamefont {McChesney}, \citenamefont {Lotsch}, \citenamefont {Schoop},\
  \ and\ \citenamefont {Ast}}]{Topp2017}%
  \BibitemOpen
  \bibfield  {author} {\bibinfo {author} {\bibfnamefont {A.} \bibnamefont
  {Topp}}, \bibinfo {author} {\bibfnamefont {R.} \bibnamefont {Queiroz}},
  \bibinfo {author} {\bibfnamefont {A.} \bibnamefont {Gr\"uneis}}, \bibinfo
  {author} {\bibfnamefont {L.} \bibnamefont {M\"uchler}}, \bibinfo {author}
  {\bibfnamefont {A.~W.} \bibnamefont {Rost}}, \bibinfo {author}
  {\bibfnamefont {A.} \bibnamefont {Varykhalov}}, \bibinfo {author}
  {\bibfnamefont {D.} \bibnamefont {Marchenko}}, \bibinfo {author}
  {\bibfnamefont {M.} \bibnamefont {Krivenkov}}, \bibinfo {author}
  {\bibfnamefont {F.} \bibnamefont {Rodolakis}}, \bibinfo {author}
  {\bibfnamefont {J.~L.} \bibnamefont {McChesney}}, \bibinfo {author}
  {\bibfnamefont {B.~V.} \bibnamefont {Lotsch}}, \bibinfo {author}
  {\bibfnamefont {L.~M.} \bibnamefont {Schoop}}, and\ \bibinfo {author}
  {\bibfnamefont {C.~R.} \bibnamefont {Ast}},\ }\href {\doibase
  10.1103/PhysRevX.7.041073} {\bibfield  {journal} {\bibinfo  {journal} {Phys.
  Rev. X}\ }\textbf {\bibinfo {volume} {7}},\ \bibinfo {pages} {041073}
  (\bibinfo {year} {2017})}\BibitemShut {NoStop}%
\bibitem [{\citenamefont {Lou}\ \emph {et~al.}(2016)\citenamefont {Lou},
  \citenamefont {Ma}, \citenamefont {Xu}, \citenamefont {Fu}, \citenamefont
  {Kong}, \citenamefont {Shi}, \citenamefont {Richard}, \citenamefont {Weng},
  \citenamefont {Fang}, \citenamefont {Sun}, \citenamefont {Wang},
  \citenamefont {Lei}, \citenamefont {Qian}, \citenamefont {Ding}, and\
  \citenamefont {Wang}}]{Lou2016}%
  \BibitemOpen
  \bibfield  {author} {\bibinfo {author} {\bibfnamefont {R.} \bibnamefont
  {Lou}}, \bibinfo {author} {\bibfnamefont {J.-Z.} \bibnamefont {Ma}},
  \bibinfo {author} {\bibfnamefont {Q.-N.} \bibnamefont {Xu}}, \bibinfo
  {author} {\bibfnamefont {B.-B.} \bibnamefont {Fu}}, \bibinfo {author}
  {\bibfnamefont {L.-Y.} \bibnamefont {Kong}}, \bibinfo {author}
  {\bibfnamefont {Y.-G.} \bibnamefont {Shi}}, \bibinfo {author} {\bibfnamefont
  {P.} \bibnamefont {Richard}}, \bibinfo {author} {\bibfnamefont {H.-M.}\
  \bibnamefont {Weng}}, \bibinfo {author} {\bibfnamefont {Z.} \bibnamefont
  {Fang}}, \bibinfo {author} {\bibfnamefont {S.-S.} \bibnamefont {Sun}},
  \bibinfo {author} {\bibfnamefont {Q.} \bibnamefont {Wang}}, \bibinfo {author}
  {\bibfnamefont {H.-C.} \bibnamefont {Lei}}, \bibinfo {author} {\bibfnamefont
  {T.} \bibnamefont {Qian}}, \bibinfo {author} {\bibfnamefont {H.} \bibnamefont
  {Ding}}, and\ \bibinfo {author} {\bibfnamefont {S.-C.} \bibnamefont
  {Wang}},\ }\href {\doibase 10.1103/PhysRevB.93.241104} {\bibfield  {journal}
  {\bibinfo  {journal} {Phys. Rev. B}\ }\textbf {\bibinfo {volume} {93}},\
  \bibinfo {pages} {241104} (\bibinfo {year} {2016})}\BibitemShut {NoStop}%
\bibitem [{\citenamefont {Liu}\ \emph {et~al.}(2016)\citenamefont {Liu},
  \citenamefont {Hu}, \citenamefont {Cao}, \citenamefont {Zhu}, \citenamefont
  {Chuang}, \citenamefont {Graf}, \citenamefont {Adams}, \citenamefont
  {Radmanesh}, \citenamefont {Spinu}, \citenamefont {Chiorescu}, and\
  \citenamefont {Mao}}]{Liu2016}%
  \BibitemOpen
  \bibfield  {author} {\bibinfo {author} {\bibfnamefont {J.} \bibnamefont
  {Liu}}, \bibinfo {author} {\bibfnamefont {J.} \bibnamefont {Hu}}, \bibinfo
  {author} {\bibfnamefont {H.} \bibnamefont {Cao}}, \bibinfo {author}
  {\bibfnamefont {Y.} \bibnamefont {Zhu}}, \bibinfo {author} {\bibfnamefont
  {A.} \bibnamefont {Chuang}}, \bibinfo {author} {\bibfnamefont
  {D.} \bibnamefont {Graf}}, \bibinfo {author} {\bibfnamefont {D.~J.}\
  \bibnamefont {Adams}}, \bibinfo {author} {\bibfnamefont {S.~M.~A.}\
  \bibnamefont {Radmanesh}}, \bibinfo {author} {\bibfnamefont {L.} \bibnamefont
  {Spinu}}, \bibinfo {author} {\bibfnamefont {I.} \bibnamefont {Chiorescu}}, 
 and\ \bibinfo {author} {\bibfnamefont {Z.} \bibnamefont {Mao}},\ }\href
  {\doibase 10.1038/srep30525} {\bibfield  {journal} {\bibinfo  {journal} {Sci.
  Rep.}\ }\textbf {\bibinfo {volume} {6}},\ \bibinfo {pages} {30525} (\bibinfo
  {year} {2016})}\BibitemShut {NoStop}%
\bibitem [{\citenamefont {Xu}\ \emph {et~al.}(2015)\citenamefont {Xu},
  \citenamefont {Song}, \citenamefont {Nie}, \citenamefont {Weng},
  \citenamefont {Fang}, and\ \citenamefont {Dai}}]{Xu2015}%
  \BibitemOpen
  \bibfield  {author} {\bibinfo {author} {\bibfnamefont {Q.} \bibnamefont
  {Xu}}, \bibinfo {author} {\bibfnamefont {Z.} \bibnamefont {Song}}, \bibinfo
  {author} {\bibfnamefont {S.} \bibnamefont {Nie}}, \bibinfo {author}
  {\bibfnamefont {H.} \bibnamefont {Weng}}, \bibinfo {author} {\bibfnamefont
  {Z.} \bibnamefont {Fang}}, and\ \bibinfo {author} {\bibfnamefont
  {X.} \bibnamefont {Dai}},\ }\href {\doibase 10.1103/PhysRevB.92.205310}
  {\bibfield  {journal} {\bibinfo  {journal} {Phys. Rev. B}\ }\textbf {\bibinfo
  {volume} {92}},\ \bibinfo {pages} {205310} (\bibinfo {year}
  {2015})}\BibitemShut {NoStop}%
\bibitem [{\citenamefont {Su}\ \emph {et~al.}(1979)\citenamefont {Su},
  \citenamefont {Schrieffer}, and\ \citenamefont {Heeger}}]{Su1979}%
  \BibitemOpen
  \bibfield  {author} {\bibinfo {author} {\bibfnamefont {W.~P.} \bibnamefont
  {Su}}, \bibinfo {author} {\bibfnamefont {J.~R.} \bibnamefont {Schrieffer}},
  and\ \bibinfo {author} {\bibfnamefont {A.~J.} \bibnamefont {Heeger}},\
  }\href {\doibase 10.1103/PhysRevLett.42.1698} {\bibfield  {journal} {\bibinfo
   {journal} {Phys. Rev. Lett.}\ }\textbf {\bibinfo {volume} {42}},\ \bibinfo
  {pages} {1698} (\bibinfo {year} {1979})}\BibitemShut {NoStop}%
\bibitem [{\citenamefont {Su}\ \emph {et~al.}(1980)\citenamefont {Su},
  \citenamefont {Schrieffer}, and\ \citenamefont {Heeger}}]{Su1980}%
  \BibitemOpen
  \bibfield  {author} {\bibinfo {author} {\bibfnamefont {W.~P.} \bibnamefont
  {Su}}, \bibinfo {author} {\bibfnamefont {J.~R.} \bibnamefont {Schrieffer}},
 and\ \bibinfo {author} {\bibfnamefont {A.~J.} \bibnamefont {Heeger}},\
  }\href {\doibase 10.1103/PhysRevB.22.2099} {\bibfield  {journal} {\bibinfo
  {journal} {Phys. Rev. B}\ }\textbf {\bibinfo {volume} {22}},\ \bibinfo
  {pages} {2099} (\bibinfo {year} {1980})}\BibitemShut {NoStop}%
\bibitem [{\citenamefont {Heeger}\ \emph {et~al.}(1988)\citenamefont {Heeger},
  \citenamefont {Kivelson}, \citenamefont {Schrieffer}, and\ \citenamefont
  {Su}}]{Heeger1988}%
  \BibitemOpen
  \bibfield  {author} {\bibinfo {author} {\bibfnamefont {A.~J.} \bibnamefont
  {Heeger}}, \bibinfo {author} {\bibfnamefont {S.} \bibnamefont {Kivelson}},
  \bibinfo {author} {\bibfnamefont {J.~R.} \bibnamefont {Schrieffer}}, and\
  \bibinfo {author} {\bibfnamefont {W.~P.} \bibnamefont {Su}},\ }\href
  {\doibase 10.1103/RevModPhys.60.781} {\bibfield  {journal} {\bibinfo
  {journal} {Rev. Mod. Phys.}\ }\textbf {\bibinfo {volume} {60}},\ \bibinfo
  {pages} {781} (\bibinfo {year} {1988})}\BibitemShut {NoStop}%
\bibitem [{\citenamefont {Peierls}(2001)}]{Peierls2001}%
  \BibitemOpen
  \bibfield  {author} {\bibinfo {author} {\bibfnamefont {R.~E.} \bibnamefont
  {Peierls}},\ }\href {\doibase 10.1093/acprof:oso/9780198507819.001.0001}
  {\emph {\bibinfo {title} {Quantum Theory of Solids}}}\ (\bibinfo  {publisher}
  {Oxford University Press},\ \bibinfo {year} {2001})\BibitemShut {NoStop}%
\bibitem [{\citenamefont {Blaha}\ \emph {et~al.}(2001)\citenamefont {Blaha},
  \citenamefont {Schwarz}, \citenamefont {Madsen}, \citenamefont {Kvasnicka},\
  \ and\ \citenamefont {Luitz}}]{P.BlahaK.SchwarzG.K.H.Madsen2001}%
  \BibitemOpen
  \bibfield  {author} {\bibinfo {author} {\bibfnamefont {P.} \bibnamefont
  {Blaha}}, \bibinfo {author} {\bibfnamefont {K.} \bibnamefont {Schwarz}},
  \bibinfo {author} {\bibfnamefont {G.~K.~H.} \bibnamefont {Madsen}}, \bibinfo
  {author} {\bibfnamefont {D.} \bibnamefont {Kvasnicka}}, and\ \bibinfo
  {author} {\bibfnamefont {J.} \bibnamefont {Luitz}},\ }\href@noop {} {\emph
  {\bibinfo {title} {{WIEN2K}, {A}n {A}ugmented {P}lane {W}ave + {L}ocal
  {O}rbitals {P}rogram for {C}alculating {C}rystal {P}roperties}}}\ (\bibinfo
  {publisher} {{K}arlheinz Schwarz, Techn. Universit\"{a}t Wien, Austria},\
  \bibinfo {year} {2001})\BibitemShut {NoStop}%
\bibitem [{\citenamefont {Becke}\ and\ \citenamefont
  {Johnson}(2006)}]{Becke2006}%
  \BibitemOpen
  \bibfield  {author} {\bibinfo {author} {\bibfnamefont {A.~D.} \bibnamefont
  {Becke}}\ and\ \bibinfo {author} {\bibfnamefont {E.~R.} \bibnamefont
  {Johnson}},\ }\href {\doibase 10.1063/1.2213970} {\bibfield  {journal}
  {\bibinfo  {journal} {J. Chem. Phys.}\ }\textbf {\bibinfo {volume} {124}},\
  \bibinfo {pages} {221101} (\bibinfo {year} {2006})}\BibitemShut {NoStop}%
\bibitem [{\citenamefont {Tran}\ and\ \citenamefont {Blaha}(2009)}]{Tran2009}%
  \BibitemOpen
  \bibfield  {author} {\bibinfo {author} {\bibfnamefont {F.} \bibnamefont
  {Tran}}\ and\ \bibinfo {author} {\bibfnamefont {P.} \bibnamefont {Blaha}},\
  }\href {\doibase 10.1103/PhysRevLett.102.226401} {\bibfield  {journal}
  {\bibinfo  {journal} {Phys. Rev. Lett.}\ }\textbf {\bibinfo {volume} {102}},\
  \bibinfo {pages} {226401} (\bibinfo {year} {2009})}\BibitemShut {NoStop}%
\bibitem [{\citenamefont {Koller}\ \emph {et~al.}(2011)\citenamefont {Koller},
  \citenamefont {Tran}, and\ \citenamefont {Blaha}}]{Koller2011}%
  \BibitemOpen
  \bibfield  {author} {\bibinfo {author} {\bibfnamefont {D.} \bibnamefont
  {Koller}}, \bibinfo {author} {\bibfnamefont {F.} \bibnamefont {Tran}}, and\
  \bibinfo {author} {\bibfnamefont {P.} \bibnamefont {Blaha}},\ }\href
  {\doibase 10.1103/PhysRevB.83.195134} {\bibfield  {journal} {\bibinfo
  {journal} {Phys. Rev. B}\ }\textbf {\bibinfo {volume} {83}},\ \bibinfo
  {pages} {195134} (\bibinfo {year} {2011})}\BibitemShut {NoStop}%
\bibitem [{\citenamefont {Li}\ \emph {et~al.}(2014)\citenamefont {Li},
  \citenamefont {Li}, \citenamefont {Blaha}, and\ \citenamefont
  {Kioussis}}]{Li2014}%
  \BibitemOpen
  \bibfield  {author} {\bibinfo {author} {\bibfnamefont {Z.} \bibnamefont
  {Li}}, \bibinfo {author} {\bibfnamefont {J.} \bibnamefont {Li}}, \bibinfo
  {author} {\bibfnamefont {P.} \bibnamefont {Blaha}}, and\ \bibinfo {author}
  {\bibfnamefont {N.} \bibnamefont {Kioussis}},\ }\href {\doibase
  10.1103/PhysRevB.89.121117} {\bibfield  {journal} {\bibinfo  {journal} {Phys.
  Rev. B}\ }\textbf {\bibinfo {volume} {89}},\ \bibinfo {pages} {121117}
  (\bibinfo {year} {2014})}\BibitemShut {NoStop}%
\bibitem [{\citenamefont {Kang}\ \emph {et~al.}(2016)\citenamefont {Kang},
  \citenamefont {Denlinger}, \citenamefont {Allen}, \citenamefont {Min},
  \citenamefont {Reinert}, \citenamefont {Kang}, \citenamefont {Cho},
  \citenamefont {Kang}, \citenamefont {Shim}, and\ \citenamefont
  {Min}}]{Kang2016}%
  \BibitemOpen
  \bibfield  {author} {\bibinfo {author} {\bibfnamefont {C.-J.} \bibnamefont
  {Kang}}, \bibinfo {author} {\bibfnamefont {J.~D.} \bibnamefont {Denlinger}},
  \bibinfo {author} {\bibfnamefont {J.~W.} \bibnamefont {Allen}}, \bibinfo
  {author} {\bibfnamefont {C.-H.} \bibnamefont {Min}}, \bibinfo {author}
  {\bibfnamefont {F.} \bibnamefont {Reinert}}, \bibinfo {author} {\bibfnamefont
  {B.~Y.} \bibnamefont {Kang}}, \bibinfo {author} {\bibfnamefont {B.~K.}\
  \bibnamefont {Cho}}, \bibinfo {author} {\bibfnamefont {J.-S.} \bibnamefont
  {Kang}}, \bibinfo {author} {\bibfnamefont {J.~H.} \bibnamefont {Shim}}, \
  \ and\ \bibinfo {author} {\bibfnamefont {B.~I.} \bibnamefont {Min}},\ }\href
  {\doibase 10.1103/PhysRevLett.116.116401} {\bibfield  {journal} {\bibinfo
  {journal} {Phys. Rev. Lett.}\ }\textbf {\bibinfo {volume} {116}},\ \bibinfo
  {pages} {116401} (\bibinfo {year} {2016})}\BibitemShut {NoStop}%
\bibitem [{\citenamefont {Lee}\ \emph {et~al.}(2013)\citenamefont {Lee},
  \citenamefont {Farhan}, \citenamefont {Kim}, and\ \citenamefont
  {Shim}}]{Lee2013}%
  \BibitemOpen
  \bibfield  {author} {\bibinfo {author} {\bibfnamefont {G.} \bibnamefont
  {Lee}}, \bibinfo {author} {\bibfnamefont {M.~A.} \bibnamefont {Farhan}},
  \bibinfo {author} {\bibfnamefont {J.~S.} \bibnamefont {Kim}}, and\
  \bibinfo {author} {\bibfnamefont {J.~H.} \bibnamefont {Shim}},\ }\href
  {\doibase 10.1103/PhysRevB.87.245104} {\bibfield  {journal} {\bibinfo
  {journal} {Phys. Rev. B}\ }\textbf {\bibinfo {volume} {87}},\ \bibinfo
  {pages} {245104} (\bibinfo {year} {2013})}\BibitemShut {NoStop}%
\bibitem [{\citenamefont {Anderson}(1990)}]{Anderson1990}%
  \BibitemOpen
  \bibfield  {author} {\bibinfo {author} {\bibfnamefont {P.~W.} \bibnamefont
  {Anderson}},\ }\href {\doibase 10.1103/PhysRevLett.64.1839} {\bibfield
  {journal} {\bibinfo  {journal} {Phys. Rev. Lett.}\ }\textbf {\bibinfo
  {volume} {64}},\ \bibinfo {pages} {1839} (\bibinfo {year}
  {1990})}\BibitemShut {NoStop}%
\bibitem [{\citenamefont {Maier}\ \emph {et~al.}(2013)\citenamefont {Maier},
  \citenamefont {Honerkamp}, and\ \citenamefont {Wang}}]{Maier2013}%
  \BibitemOpen
  \bibfield  {author} {\bibinfo {author} {\bibfnamefont {S.~A.} \bibnamefont
  {Maier}}, \bibinfo {author} {\bibfnamefont {C.} \bibnamefont {Honerkamp}}, \
  and\ \bibinfo {author} {\bibfnamefont {Q.-H.} \bibnamefont {Wang}},\ }\href
  {\doibase 10.3390/sym5040313} {\bibfield  {journal} {\bibinfo  {journal}
  {Symmetry}\ }\textbf {\bibinfo {volume} {5}},\ \bibinfo {pages} {313}
  (\bibinfo {year} {2013})}\BibitemShut {NoStop}%
\bibitem [{\citenamefont {Bena}\ and\ \citenamefont
  {Montambaux}(2009)}]{Bena2009}%
  \BibitemOpen
  \bibfield  {author} {\bibinfo {author} {\bibfnamefont {C.} \bibnamefont
  {Bena}}\ and\ \bibinfo {author} {\bibfnamefont {G.} \bibnamefont
  {Montambaux}},\ }\href {http://stacks.iop.org/1367-2630/11/i=9/a=095003}
  {\bibfield  {journal} {\bibinfo  {journal} {New J. Phys.}\ }\textbf {\bibinfo
  {volume} {11}},\ \bibinfo {pages} {095003} (\bibinfo {year}
  {2009})}\BibitemShut {NoStop}%
\bibitem [{\citenamefont {Kochan}\ \emph {et~al.}(2017)\citenamefont {Kochan},
  \citenamefont {Irmer}, and\ \citenamefont {Fabian}}]{Kochan2017}%
  \BibitemOpen
  \bibfield  {author} {\bibinfo {author} {\bibfnamefont {D.} \bibnamefont
  {Kochan}}, \bibinfo {author} {\bibfnamefont {S.} \bibnamefont {Irmer}}, 
and\ \bibinfo {author} {\bibfnamefont {J.} \bibnamefont {Fabian}},\ }\href
  {\doibase 10.1103/PhysRevB.95.165415} {\bibfield  {journal} {\bibinfo
  {journal} {Phys. Rev. B}\ }\textbf {\bibinfo {volume} {95}},\ \bibinfo
  {pages} {165415} (\bibinfo {year} {2017})}\BibitemShut {NoStop}%
\bibitem [{\citenamefont {Huang}\ \emph {et~al.}(2017)\citenamefont {Huang},
  \citenamefont {Kim}, \citenamefont {Shelton}, \citenamefont {Plummer}, and\
  \citenamefont {Jin}}]{Huang2017}%
  \BibitemOpen
  \bibfield  {author} {\bibinfo {author} {\bibfnamefont {S.} \bibnamefont
  {Huang}}, \bibinfo {author} {\bibfnamefont {J.} \bibnamefont {Kim}}, \bibinfo
  {author} {\bibfnamefont {W.~A.} \bibnamefont {Shelton}}, \bibinfo {author}
  {\bibfnamefont {E.~W.} \bibnamefont {Plummer}}, and\ \bibinfo {author}
  {\bibfnamefont {R.} \bibnamefont {Jin}},\ }\href {\doibase
  10.1073/pnas.1706657114} {\bibfield  {journal} {\bibinfo  {journal} {Proc.
  Natl Acad. Sci. USA}\ }\textbf {\bibinfo {volume} {114}},\ \bibinfo {pages}
  {6256} (\bibinfo {year} {2017})}\BibitemShut {NoStop}%
\bibitem [{\citenamefont {Luo}\ and\ \citenamefont {Xiang}(2015)}]{Luo2015}%
  \BibitemOpen
  \bibfield  {author} {\bibinfo {author} {\bibfnamefont {W.} \bibnamefont
  {Luo}}\ and\ \bibinfo {author} {\bibfnamefont {H.} \bibnamefont {Xiang}},\
  }\href {\doibase 10.1021/acs.nanolett.5b00418} {\bibfield  {journal}
  {\bibinfo  {journal} {Nano Lett.}\ }\textbf {\bibinfo {volume} {15}},\
  \bibinfo {pages} {3230} (\bibinfo {year} {2015})}\BibitemShut {NoStop}%
\bibitem [{\citenamefont {He}\ \emph {et~al.}(2017)\citenamefont {He},
  \citenamefont {Fu}, \citenamefont {Zhao}, \citenamefont {Liang},
  \citenamefont {Chen}, \citenamefont {Leng}, \citenamefont {Wang},
  \citenamefont {Li}, \citenamefont {Zhang}, \citenamefont {Xue}, \citenamefont
  {Li}, \citenamefont {Zhang}, \citenamefont {Ren}, and\ \citenamefont
  {Chen}}]{He2017b}%
  \BibitemOpen
  \bibfield  {author} {\bibinfo {author} {\bibfnamefont {J.~B.} \bibnamefont
  {He}}, \bibinfo {author} {\bibfnamefont {Y.} \bibnamefont {Fu}}, \bibinfo
  {author} {\bibfnamefont {L.~X.} \bibnamefont {Zhao}}, \bibinfo {author}
  {\bibfnamefont {H.} \bibnamefont {Liang}}, \bibinfo {author} {\bibfnamefont
  {D.} \bibnamefont {Chen}}, \bibinfo {author} {\bibfnamefont {Y.~M.}\
  \bibnamefont {Leng}}, \bibinfo {author} {\bibfnamefont {X.~M.} \bibnamefont
  {Wang}}, \bibinfo {author} {\bibfnamefont {J.} \bibnamefont {Li}}, \bibinfo
  {author} {\bibfnamefont {S.} \bibnamefont {Zhang}}, \bibinfo {author}
  {\bibfnamefont {M.~Q.} \bibnamefont {Xue}}, \bibinfo {author} {\bibfnamefont
  {C.~H.} \bibnamefont {Li}}, \bibinfo {author} {\bibfnamefont
  {P.} \bibnamefont {Zhang}}, \bibinfo {author} {\bibfnamefont {Z.~A.}\
  \bibnamefont {Ren}}, and\ \bibinfo {author} {\bibfnamefont {G.~F.}\
  \bibnamefont {Chen}},\ }\href {\doibase 10.1103/PhysRevB.95.045128}
  {\bibfield  {journal} {\bibinfo  {journal} {Phys. Rev. B}\ }\textbf {\bibinfo
  {volume} {95}},\ \bibinfo {pages} {045128} (\bibinfo {year}
  {2017})}\BibitemShut {NoStop}%
\bibitem [{\citenamefont {Farhan}\ \emph {et~al.}(2014)\citenamefont {Farhan},
  \citenamefont {Lee}, and\ \citenamefont {Shim}}]{Farhan2014}%
  \BibitemOpen
  \bibfield  {author} {\bibinfo {author} {\bibfnamefont {M.~A.} \bibnamefont
  {Farhan}}, \bibinfo {author} {\bibfnamefont {G.} \bibnamefont {Lee}}, and\
  \bibinfo {author} {\bibfnamefont {J.~H.} \bibnamefont {Shim}},\ }\href
  {http://stacks.iop.org/0953-8984/26/i=4/a=042201} {\bibfield  {journal}
  {\bibinfo  {journal} {J. Phys.: Condens. Matter}\ }\textbf {\bibinfo {volume}
  {26}},\ \bibinfo {pages} {042201} (\bibinfo {year} {2014})}\BibitemShut
  {NoStop}%
\bibitem [{\citenamefont {Kitaev}(2009)}]{Kitaev2009}%
  \BibitemOpen
  \bibfield  {author} {\bibinfo {author} {\bibfnamefont {A.} \bibnamefont
  {Kitaev}},\ }\href {\doibase 10.1063/1.3149495} {\bibfield  {journal}
  {\bibinfo  {journal} {AIP Conf. Proc.}\ }\textbf {\bibinfo {volume} {1134}},\
  \bibinfo {pages} {22} (\bibinfo {year} {2009})}\BibitemShut {NoStop}%
\bibitem [{\citenamefont {Ryu}\ \emph {et~al.}(2010)\citenamefont {Ryu},
  \citenamefont {Schnyder}, \citenamefont {Furusaki}, and\ \citenamefont
  {Ludwig}}]{Ryu2010}%
  \BibitemOpen
  \bibfield  {author} {\bibinfo {author} {\bibfnamefont {S.} \bibnamefont
  {Ryu}}, \bibinfo {author} {\bibfnamefont {A.~P.} \bibnamefont {Schnyder}},
  \bibinfo {author} {\bibfnamefont {A.} \bibnamefont {Furusaki}}, and\
  \bibinfo {author} {\bibfnamefont {A.~W.~W.} \bibnamefont {Ludwig}},\ }\href
  {http://stacks.iop.org/1367-2630/12/i=6/a=065010} {\bibfield  {journal}
  {\bibinfo  {journal} {New J. Phys.}\ }\textbf {\bibinfo {volume} {12}},\
  \bibinfo {pages} {065010} (\bibinfo {year} {2010})}\BibitemShut {NoStop}%
\bibitem [{\citenamefont {Lee}\ and\ \citenamefont {Fabian}(2016)}]{Lee2016}%
  \BibitemOpen
  \bibfield  {author} {\bibinfo {author} {\bibfnamefont {J.} \bibnamefont
  {Lee}}\ and\ \bibinfo {author} {\bibfnamefont {J.} \bibnamefont {Fabian}},\
  }\href {\doibase 10.1103/PhysRevB.94.195401} {\bibfield  {journal} {\bibinfo
  {journal} {Phys. Rev. B}\ }\textbf {\bibinfo {volume} {94}},\ \bibinfo
  {pages} {195401} (\bibinfo {year} {2016})}\BibitemShut {NoStop}%
\bibitem [{\citenamefont {Kane}\ and\ \citenamefont {Mele}(2005)}]{Kane2005a}%
  \BibitemOpen
  \bibfield  {author} {\bibinfo {author} {\bibfnamefont {C.~L.} \bibnamefont
  {Kane}}\ and\ \bibinfo {author} {\bibfnamefont {E.~J.} \bibnamefont
  {Mele}},\ }\href {\doibase 10.1103/PhysRevLett.95.146802} {\bibfield
  {journal} {\bibinfo  {journal} {Phys. Rev. Lett.}\ }\textbf {\bibinfo
  {volume} {95}},\ \bibinfo {pages} {146802} (\bibinfo {year}
  {2005})}\BibitemShut {NoStop}%
\bibitem [{\citenamefont {Min}\ \emph {et~al.}(2006)\citenamefont {Min},
  \citenamefont {Hill}, \citenamefont {Sinitsyn}, \citenamefont {Sahu},
  \citenamefont {Kleinman}, and\ \citenamefont {MacDonald}}]{Min2006}%
  \BibitemOpen
  \bibfield  {author} {\bibinfo {author} {\bibfnamefont {H.} \bibnamefont
  {Min}}, \bibinfo {author} {\bibfnamefont {J.~E.} \bibnamefont {Hill}},
  \bibinfo {author} {\bibfnamefont {N.~A.} \bibnamefont {Sinitsyn}}, \bibinfo
  {author} {\bibfnamefont {B.~R.} \bibnamefont {Sahu}}, \bibinfo {author}
  {\bibfnamefont {L.} \bibnamefont {Kleinman}}, and\ \bibinfo {author}
  {\bibfnamefont {A.~H.} \bibnamefont {MacDonald}},\ }\href {\doibase
  10.1103/PhysRevB.74.165310} {\bibfield  {journal} {\bibinfo  {journal} {Phys.
  Rev. B}\ }\textbf {\bibinfo {volume} {74}},\ \bibinfo {pages} {165310}
  (\bibinfo {year} {2006})}\BibitemShut {NoStop}%
\bibitem [{\citenamefont {Liu}\ \emph {et~al.}(2011)\citenamefont {Liu},
  \citenamefont {Jiang}, and\ \citenamefont {Yao}}]{Liu2011b}%
  \BibitemOpen
  \bibfield  {author} {\bibinfo {author} {\bibfnamefont {C.-C.} \bibnamefont
  {Liu}}, \bibinfo {author} {\bibfnamefont {H.} \bibnamefont {Jiang}}, and\
  \bibinfo {author} {\bibfnamefont {Y.} \bibnamefont {Yao}},\ }\href {\doibase
  10.1103/PhysRevB.84.195430} {\bibfield  {journal} {\bibinfo  {journal} {Phys.
  Rev. B}\ }\textbf {\bibinfo {volume} {84}},\ \bibinfo {pages} {195430}
  (\bibinfo {year} {2011})}\BibitemShut {NoStop}%
\bibitem [{\citenamefont {Li}\ \emph {et~al.}(2016)\citenamefont {Li},
  \citenamefont {Wang}, \citenamefont {Graf}, \citenamefont {Wang},
  \citenamefont {Wang}, and\ \citenamefont {Petrovic}}]{Li2016}%
  \BibitemOpen
  \bibfield  {author} {\bibinfo {author} {\bibfnamefont {L.} \bibnamefont
  {Li}}, \bibinfo {author} {\bibfnamefont {K.} \bibnamefont {Wang}}, \bibinfo
  {author} {\bibfnamefont {D.} \bibnamefont {Graf}}, \bibinfo {author}
  {\bibfnamefont {L.} \bibnamefont {Wang}}, \bibinfo {author} {\bibfnamefont
  {A.} \bibnamefont {Wang}}, and\ \bibinfo {author} {\bibfnamefont
  {C.} \bibnamefont {Petrovic}},\ }\href {\doibase 10.1103/PhysRevB.93.115141}
  {\bibfield  {journal} {\bibinfo  {journal} {Phys. Rev. B}\ }\textbf {\bibinfo
  {volume} {93}},\ \bibinfo {pages} {115141} (\bibinfo {year}
  {2016})}\BibitemShut {NoStop}%
\bibitem [{\citenamefont {Lee}()}]{Inho}%
  \BibitemOpen
  \bibfield  {author} {\bibinfo {author} {\bibfnamefont {I.} \bibnamefont
  {Lee}},\ }\href@noop {} {}\bibinfo {howpublished} {to be
  published}\BibitemShut {NoStop}%
\end{thebibliography}
%



\let\origdescription\description
\renewenvironment{description}{
\setlength{\leftmargini}{0em}
\origdescription
\setlength{\itemindent}{0em}
\setlength{\labelsep}{\textwidth}
}
{\endlist}


\end{document}